\documentclass[3p,12pt]{elsarticle}
\pdfoutput=1
\usepackage{amssymb,amsmath,graphicx,hyperref,color,bm}

\numberwithin{equation}{section}

\graphicspath{{fig/}}

\DeclareMathOperator{\Tr}{Tr}

\journal{Nuclear Physics B}

\makeatletter
\def\@author#1{\g@addto@macro\elsauthors{\normalsize%
    \def\baselinestretch{1}%
    \upshape\authorsep#1\unskip\textsuperscript{%
      \ifx\@fnmark\@empty\else\unskip\sep\@fnmark\let\sep=,\fi
      \ifx\@corref\@empty\else\unskip\sep\@corref\let\sep=,\fi
      }%
    \def\authorsep{\space {\large and}\space}%
    \global\let\@fnmark\@empty
    \global\let\@corref\@empty
    \global\let\sep\@empty}%
    \@eadauthor={#1}
}

\def\@@author[#1]#2{\g@addto@macro\elsauthors{%
    \def\baselinestretch{1}%
    \authorsep#2\unskip\textsuperscript{
      \@for\@@affmark:=#1\do{%
       \edef\affnum{\@ifundefined{X@\@@affmark}{1}{\elsRef{\@@affmark}}}%
     \unskip\sep\affnum\let\sep=,}%
      \ifx\@fnmark\@empty\else\unskip\sep\@fnmark\let\sep=,\fi
      \ifx\@corref\@empty\else\unskip\sep\@corref\let\sep=,\fi
      }%
    \def\authorsep{\space {\large and}\space}%
    \global\let\sep\@empty\global\let\@corref\@empty
    \global\let\@fnmark\@empty}%
    \@eadauthor={#2}
}

\def\ps@pprintTitle{%
 \let\@oddhead\@empty
 \let\@evenhead\@empty
 \def\@oddfoot{}%
 \let\@evenfoot\@oddfoot}

\long\def\MaketitleBox{%
  \resetTitleCounters
  \def\baselinestretch{1}%
  \begin{center}%
   \def\baselinestretch{1}%
    \Large\@title\par\vskip18pt
    \normalsize\elsauthors\par\vskip10pt
    \footnotesize\itshape\elsaddress\par\vskip36pt
    \rule{\textwidth}{1.5pt}\vskip12pt
    \ifvoid\absbox\else\unvbox\absbox\par\vskip10pt\fi
    \ifvoid\keybox\else\unvbox\keybox\par\vskip10pt\fi
    \rule{\textwidth}{1.5pt}\vskip12pt
    \end{center}%
  }

\renewcommand\subsection{\@startsection{subsection}{2}{\z@}%
           {18\p@ \@plus 6\p@ \@minus 3\p@}%
           {9\p@ \@plus 6\p@ \@minus 3\p@}%
           {\normalfont\normalsize\itshape\bfseries}}

\gdef\emailauthor#1#2{\stepcounter{ead}%
     \g@addto@macro\@elseads{\raggedright%
      \let\corref\@gobble
      \eadsep\newline\texttt{#1} (#2)\def\eadsep{\unskip,\space}}%
}

\renewcommand\appendix{\par
  \setcounter{section}{0}%
  \setcounter{subsection}{0}%
  \setcounter{equation}{0}
  \gdef\thefigure{\arabic{figure}}%
  \gdef\thetable{\@Alph\c@section.\arabic{table}}%
  \gdef\thesection{\appendixname~\@Alph\c@section}%
  \@addtoreset{equation}{section}%
  \gdef\theequation{\@Alph\c@section.\arabic{equation}}%
  \addtocontents{toc}{\string\let\string\numberline\string\tmptocnumberline}{}{}
}

\def\appendixname{Appendix}
\renewcommand\@makefntext[1]{#1}

\makeatother

\biboptions{numbers,sort&compress}

\newcommand{\TCT}{\widetilde{\bm{\mathcal{T}}}_{\!\!\rm CT}}
\newcommand{\ICT}{\widetilde{\bm{\mathcal{I}}}_{\rm CT}}
\newcommand{\JCT}{\widetilde{\bm{\mathcal{J}}}_{\!\rm CT}}

\begin{document}

\begin{frontmatter}

${}$
\vspace{-2cm}
\begin{flushright}
MAN/HEP/2015/18,
ULB-TH/15-19\\
November 2015
\end{flushright}
\medskip
\title{{\bf {\LARGE Symmetry-Improved 2PI Approach}}\\[2mm] 
{\bf {\LARGE to the Goldstone-Boson IR Problem}}\\[2mm] 
{\bf {\LARGE of the SM Effective Potential}}\bigskip}

\author[uom]{{\large Apostolos Pilaftsis}}
\author[uom,ulb]{{\large Daniele Teresi}}

\address[uom]{\smallskip Consortium for Fundamental Physics, School of
  Physics and Astronomy,\\ University of Manchester, Manchester M13
  9PL, United Kingdom\\[1mm]
{\rm E-mail address:} {\tt apostolos.pilaftsis@manchester.ac.uk}\bigskip} 
  
\address[ulb]{\smallskip  Service de Physique Th\'eorique, Universit\'e Libre de Bruxelles,\\
Boulevard du Triomphe, CP225, 1050 Brussels, Belgium\\[1mm]
{\rm E-mail address:} {\tt daniele.teresi@ulb.ac.be}}

\begin{abstract}
The effective potential of the Standard Model~(SM), from three loop order and higher, suffers from infra\-red~(IR) divergences arising from quantum effects due to massless would-be Goldstone bosons associated with the longitudinal polarizations of the $W^\pm$ and $Z$ bosons. Such IR pathologies also hinder accurate evaluation of the two-loop threshold corrections to electroweak quantities, such as the vacuum expectation value of the Higgs field. However, these divergences are an artifact of perturbation theory, and therefore need to be consistently resummed in order to obtain an IR-safe effective potential. The so-called Two-Particle-Irreducible (2PI) effective action provides a rigorous framework to consistently perform such resummations, without the need to resort to {\it ad hoc} subtractions or running into the risk of over-counting contributions. By considering the recently proposed symmetry-improved 2PI formalism, we address the problem of the Goldstone-boson IR divergences of the SM effective potential in the gaugeless limit of the theory. In the same limit, we evaluate the IR-safe symmetry-improved 2PI effective potential, after taking into account quantum loops of chiral fermions, as well as the renormalization of spurious custodially breaking effects triggered by fermionic Yukawa interactions. Finally, we compare our results with those obtained with other methods presented in the literature.

\bigskip

\end{abstract}


\end{frontmatter}

\vfill\eject

\makeatletter
\renewcommand\@makefntext[1]{\leftskip=0em\hskip1em\@makefnmark\space #1}
\makeatother

\section{Introduction}

In Quantum Field Theory (QFT), there are instances where fixed-order perturbative expansions break down and one needs to rely on techniques for resumming higher-order contributions to deal with this problem. A few typical examples are: the IR problem in thermal QFT at high temperatures~\cite{Weinberg:1974hy, Dolan:1973qd, Blaizot:2000fc, Berges:2004hn}, the problem of pinch singularities in non-equilibrium QFT~\cite{Berges:2004yj,Millington:2012pf}, the dynamical generation of an effective gluon mass~\cite{Aguilar:2008xm,Aguilar:2011ux}, the resonant production and mixing of unstable particles~\cite{Pilaftsis:1989zt,Pilaftsis:1997dr}, as the latter occurs, for example, in scenarios of resonant leptogenesis~\cite{Pilaftsis:1997jf, Pilaftsis:2003gt, Garny:2009qn, Garbrecht:2011aw, Garny:2011hg, Dev:2014laa, Dev:2014wsa}. On the other hand, there are cases in which higher-order effects could play an important role, even in scenarios with small perturbative couplings and non-resonant dynamics.  For instance, recent studies~\cite{Bezrukov:2012sa, Degrassi:2012ry, Buttazzo:2013uya} indicate that the profile of the SM effective potential, extra\-polated to very high energies, is extremely sensitive to the physics at the electroweak scale. Thus, a formalism to incorporate and resum higher-order effects in a rigorous and self-consistent manner is highly desirable for a number of applications in thermal and
non-thermal QFT.

Recently, it was pointed out~\cite{Martin:2013gka} that the conventional One-Particle-Irreducible~(1PI) effective potential~\cite{Goldstone:1962es, JonaLasinio:1964cw, Coleman:1973jx} of the SM is plagued by IR divergences caused by quantum effects due to {\em massless} would-be Goldstone bosons associated with the longitudinal polarizations of the $W^\pm$ and $Z$ bosons. These divergences start from three-loop order for the effective potential~$V_{\rm eff}(\phi )$ itself, but from two loops for its derivative with respect to the Higgs background field~$\phi$, $dV_{\rm eff}(\phi)/d\phi$, which is required for determining the vacuum expectation value~(VEV) of~$\phi$.  The latter is a key quantity, as it enters the state-of-the-art calculations of the matching conditions for the SM effective potential at the electroweak scale.

The IR divergences in the 1PI effective potential pose a serious field-theoretic problem which needs to be addressed for two reasons. First, we expect conceptually that the effective potential $V_{\rm eff}(\phi )$ is a well-behaved analytic function for all values of~$\phi$. Second, we observe that these IR pathologies formally lower the loop order of the involved contributions, thus causing a breakdown of perturbation theory. Therefore, loop graphs that are naively of higher order can potentially give significant contributions to the threshold corrections to the VEV of $\phi$. Since the precise functional form of the effective potential~$V_{\rm eff} (\phi )$ for high values of $\phi$ is very sensitive to the matching conditions at the electroweak scale, this IR problem may affect the stability analyses of the SM potential. Most recently, this problem was addressed~\cite{Martin:2014bca, Elias-Miro:2014pca} by devising a procedure for resumming the pathological IR-divergent terms to all orders, albeit in an {\it ad hoc} manner.

A rigorous framework to study the IR problem of the 1PI effective potential is the formalism introduced by Cornwall, Jackiw and Tomboulis~(CJT)~\cite{Cornwall:1974vz}.  In its simplest version, the Two-Particle-Irreducible (2PI) effective action is a generating functional expressed not only in terms of fields, but also in terms of their dressed propagators. At any given order of its loopwise expansion, the 2PI effective action represents an infinite set of higher-order diagrams induced by partially resummed propagators. Most importantly, in this 2PI approach of selective resummations, one does not run into the danger of over-counting graphs.

There have been numerous applications of the 2PI formalism in the literature, although the main focus of these were within the context of thermal QFT~\cite{AmelinoCamelia:1992nc, Petropoulos:1998gt, Blaizot:2000fc,Berges:2004hn, Reinosa:2011ut,Marko:2013lxa, Marko:2015gpa}. Nevertheless, one major limitation of the CJT formalism remains its difficulty to describe properly the global and local symmetries of the theory, at any fixed order of a loopwise expansion of the 2PI effective action. In particular, in the case of global symmetries, higher-order effects distort them at any given order of the loopwise expansion, giving rise to massive Goldstone bosons in the Spontaneous Symmetry Breaking (SSB) phase of the theory~\cite{Baym:1977qb, AmelinoCamelia:1997dd, Petropoulos:1998gt}. A~satisfactory solution to this problem may be obtained within the context of the recently proposed \emph{symmetry-improved CJT formalism}~\cite{Pilaftsis:2013xna}.  In this formalism, the effective potential is defined by virtue of the 1PI Ward Identity (WI) associated with the global symmetry. This Symmetry-Improved Two-Particle-Irreducible~(SI2PI) approach has a number of desirable field-theoretical properties~\cite{Pilaftsis:2013xna} that ensure the masslessness of the Goldstone bosons within quantum loops. Recently, the SI2PI formalism has also been used to study the chiral phase transition~\cite{Mao:2013gva} in an $O(4)$ theory, and it has been extended to higher $n$PI effective actions~\cite{Brown:2015xma}. In the same context, possible alternative 2PI formulations~\cite{Garbrecht:2015cla} have been suggested.

In this paper we calculate the SM effective potential with chiral fermions in the gaugeless limit of theory, within the SI2PI formalism. Specifically, we consider an {\it ungauged} model based on the $SU(2)_L \times U(1)_Y$ group with one Higgs doublet, one doublet of left-handed top and bottom quarks, and one right-handed top quark. Moreover, quantum effects due to chiral fermions are treated semi-perturbatively, in the sense that the double Legendre transform giving rise to the 2PI effective action is performed only with respect to the scalar fields. We expect that these approximations yield a relatively accurate evaluation of the full SM effective potential.  In this simplified framework of the SI2PI formalism, we study the problem of IR divergences of the SM effective potential.  In particular, we compute the SI2PI effective potential and show that it is IR safe. For comparison, we first consider only the scalar-boson contributions, by neglecting fermion quantum effects. We find that our results differ in a relevant manner with those obtained using the approximate partial resummation method of~\cite{Martin:2014bca, Elias-Miro:2014pca}, thereby confirming the preliminary analysis given in~\cite{Pilaftsis:2015cka}. Then, we include the contributions from quantum fermion loops to find that our results are in fair agreement with those reported in~\cite{Martin:2014bca,Elias-Miro:2014pca}.

The layout of the paper is as follows. After this introductory section, in Section~\ref{sec:IR} we discuss how the IR divergences appear in the 1PI effective potential. Also, we briefly present an approximate partial resummation prescription that enables one to deal with the Goldstone-boson IR problem. In Section~\ref{sec:scalar} we review the SI2PI formalism, which we apply to the SM scalar sector, based on the SSB of the global $SU(2)_L \times U(1)_Y$ group. In~Section~\ref{sec:ferm} we include the contribution from chiral fermion quantum loops, specifically due to top and bottom quarks. In addition, we describe our renormalization programme of the SI2PI effective action, which includes renormalization of spurious custodially breaking effects triggered by Yukawa interactions.  In Section~\ref{sec:IR_2PI} we compute the IR-safe SI2PI effective potential, in which both SM scalar and fermion loops are considered. We compare our numerical estimates with the ones obtained with the method of~\cite{Martin:2014bca,Elias-Miro:2014pca}. Section~\ref{sec:conclusions} presents our conclusions. Finally, pertinent technical details and detailed formulae were relegated to the two Appendices A and B.

\section{The Infrared Divergences of the SM Effective Potential}
\label{sec:IR}

In this section, we will demonstrate how quantum loops of massless Goldstone bosons can cause IR divergences in the 1PI effective potential of the SM~\cite{Martin:2013gka}. Also, we will briefly review the prescription presented in~\cite{Martin:2014bca, Elias-Miro:2014pca} to deal with this IR problem, which is based on an approximate partial resummation of a selected topology of graphs. The results obtained with this approximate method will be compared in Section~\ref{sec:IR_2PI} with those derived by employing our SI2PI approach.

\begin{figure}[t] \centering \begin{tabular}{rlrlrl}
                               $\parbox{4em}{\includegraphics[width=4em]{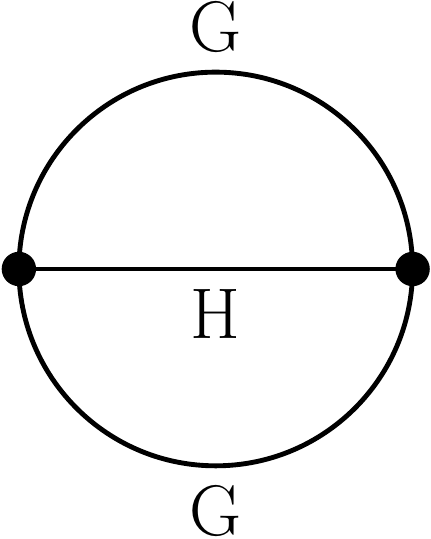}}$ & $\sim\; g \ln g$ & $\parbox{6em}{\includegraphics[width=6em]{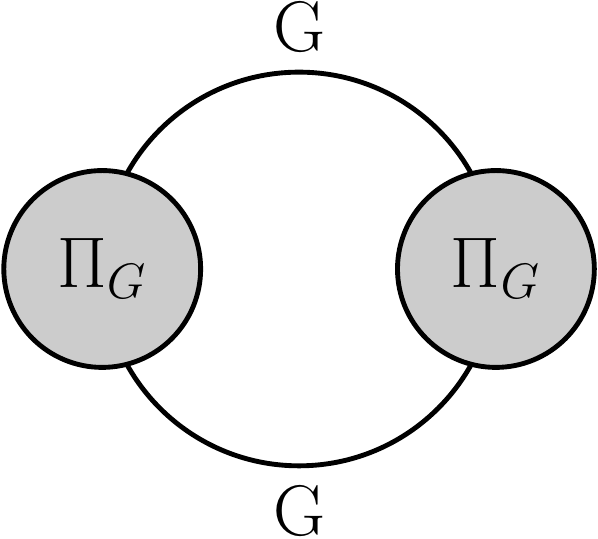}}$ & $\sim \; \ln g$ & $\parbox{6em}{\includegraphics[width=6em]{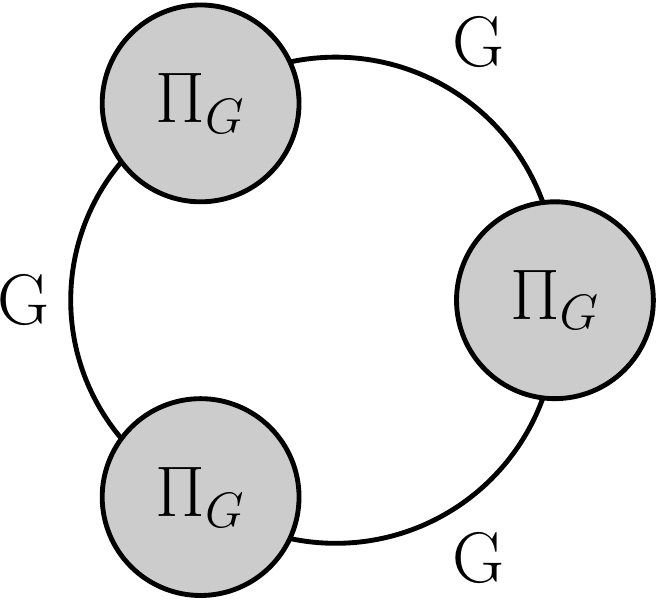}}$ & $\sim \; \displaystyle \frac{1}{g}$\\
                               \rule{0em}{4em} $\displaystyle \frac{d}{d \phi} \; \parbox{4em}{\includegraphics[width=4em]{GGH}}$ & $\sim \; \ln g$ & $\displaystyle \frac{d}{d \phi} \; \parbox{6em}{\includegraphics[width=6em]{GbGb}}$ & $\sim \; \displaystyle \frac{1}{g}$ & $\displaystyle \frac{d}{d \phi} \; \parbox{6em}{\includegraphics[width=6em]{GbGbGb}}$ & $\displaystyle \sim \;\frac{1}{g^2}$ \vspace{0.5em} 
                             \end{tabular} 
\caption{IR dependence of the Goldstone-boson ring contributions to the 1PI effective potential~$V_{\rm eff}(\phi)$  and its $\phi$-derivative $d V_{\rm eff}/ d \phi$, as a function of the squared Goldstone-boson mass $g\equiv m^2_G$. The IR infinities start from three loops for $V_{\rm eff}(\phi)$ and from two loops for $d V_{\rm eff}/ d \phi$.\label{fig:IR}}
\end{figure}

From a given order and higher in perturbation theory, the SM effective potential $V_{\rm eff} (\phi )$ suffers from IR divergences due to the presence of Goldstone bosons in ring diagrams, as shown in Figure~\ref{fig:IR}.  This IR problem starts at three-loop order, where the divergence is logarithmic, and becomes more severe with increasing loop order.  As also shown in Figure~\ref{fig:IR}, these IR divergences become even more severe when one considers the derivative of the effective potential $d V_{\rm eff}/d \phi$, in which case the IR infinities appear in two loops.

In the usual perturbation theory, the IR divergences stem from the value of the Higgs field~$\phi$, for which the neutral and charged Goldstone-boson propagators $\Delta^{G,+}(k)$ exhibit massless poles at the tree level. Evidently, this happens at the minimum of the tree-level potential, i.e.~when 
\begin{equation}
  \label{mG2}
k^2\ =\ m^2_G\ \equiv\  \lambda \phi^2 - m^2\ \to\ 0
\end{equation} 
in the Landau gauge $\xi = 0$, where $\lambda$ and $m^2$ are the quartic coupling and the squared mass term of the SM potential, respectively. Instead, the effective potential~$V_{\rm eff} (\phi )$ and all its $\phi$-derivatives are finite at the dressed minimum $\phi = v$. Nevertheless, the IR infinities in~$V_{\rm eff} (\phi )$ for some values of $\phi$ still pose a serious field-theoretic problem, for the following reasons:

\begin{itemize}

\item[(i)] The effective potential $V_{\rm eff}(\phi)$ should be a well-defined analytic function for all values of $\phi$, and not only at its dressed minimum $\phi = v$. Interestingly enough, the functional form of the effective potential at $\phi \neq v$ is an essential quantity in inflationary scenarios,
as it governs the dynamics of the background inflaton field.

\item[(ii)] At the dressed minimum $\phi = v$, the dressed Goldstone-boson masses vanish. Hence, their tree-level mass $m_G^2$ given in~\eqref{mG2} is formally of the same order as the one-loop Goldstone-boson self-energy $\Pi_G^{(1)}(k)$ at $k^2 = 0$.  As can be seen from Figure~\ref{fig:IR}, starting from three-loop order, the would-be IR infinities of $d V_{\rm eff}/d \phi$ are of the form $1/(m_G^2)^n$, with $n \geq 1$. Consequently, {\em all} higher-loop contributions to $d V_{\rm eff}/d \phi$ are formally of two-loop order. Clearly, this signifies a breakdown of perturbation theory and these diagrams can potentially have a significant impact on the two-loop evaluation for~$d V_{\rm eff}/d \phi$, and so on the state-of-the-art threshold corrections to the VEV~$v$. Because of the extreme sensitivity of the SM effective potential at high $\phi$ values to the matching conditions at the electroweak scale~\cite{Bezrukov:2012sa,Degrassi:2012ry, Buttazzo:2013uya}, it is obvious that an IR-sensitive value of $\phi$ at~$\phi = v$ may have a relevant impact on the stability analyses of the SM itself.

\end{itemize}

\vfill\eject

Another related problem is that the {\em squared} tree-level mass of the Goldstone boson $m^2_G$ can be negative at the dressed minimum $\phi = v$, thus generating an unphysical imaginary part for the SM effective potential at its minimum, which does not correspond to a true instability of the homogeneous vacuum.  This suggests that a resummation of higher-loop diagrams is needed to address this conceptual problem as well. Finally, we stress that, even though the location of the IR divergences depends on the value of the gauge-fixing parameter~$\xi$, the divergences  are nonetheless present in any renormalizable $R_\xi$ gauge at field values of $\phi$, for which $m_G^2(\phi; \xi)=0$.

\subsection{Approximate Partial Resummation}\label{sec:IR_pert}

It is now interesting to briefly outline an approximate partial resummation procedure which was proposed in~\cite{Martin:2014bca,Elias-Miro:2014pca} to address the IR problem in the 1PI effective potential.

This procedure consists in considering only ring diagrams, as displayed in Figure~\ref{fig:IR}, with insertions of one-loop Goldstone-boson self-energies $\Pi_G(k)$. In this approach, the Goldstone-boson self-energies were approximated with their zero-momentum value $\Pi_G(0)$. With this important simplification, the ring diagrams can, in principle, be resummed which results in making the following replacement for the one-loop Goldstone-boson contribution to the 1PI effective potential: 
\begin{equation} 
V_{\mathrm{eff}, G}^{(1)} \ = \ \frac{3 \, m^4_G}{4 \, (16 \pi^2)} \bigg[ \ln\bigg(\frac{m^2_G}{\mu^2} \bigg) - \frac{3}{2}\bigg] \quad \longrightarrow \quad \frac{3 \, (m^2_G + \Pi_G(0))^2}{4 \, (16 \pi^2)} \bigg[ \ln\bigg(\frac{m^2_G + \Pi_G(0)}{\mu^2} \bigg) - \frac{3}{2}\bigg]\;.  
\end{equation} 
However, as argued in \cite{Martin:2014bca, Elias-Miro:2014pca}, the $\phi$-derivative of such a resummed term is still divergent. This pathology can be remedied by prescribing that~$\Pi_G(0)$ only contains terms that are {\em not} proportional to~$m_G^2$, which amounts to replacing \begin{equation}
  \label{Pig}
\Pi_G(0) \quad \longrightarrow \quad  \Pi_g \ \equiv \ \Pi_G(0) - \frac{3 \lambda}{(16 \pi^2)} \, m^2_G \Big(\ln(m^2_G/\mu^2) - 1\Big) \;.
\end{equation}
Note that the subtracted term from $\Pi_G(0)$ in~\eqref{Pig} does not correspond to a particular diagram, rather it represents an {\it ad hoc} choice of contributions from a tadpole-like self-energy graph involving a single Goldstone-boson loop and a sunset graph with one Higgs and one Goldstone boson running in the loop. As a last step of the prescription adopted in~\cite{Martin:2014bca, Elias-Miro:2014pca}, one needs to remove from $V_{\rm eff}$  {\it by hand} all those diagrams that would be double-counted otherwise. Taking all these facts into account, the partially resummed effective potential reads:
\begin{equation}
V_{\mathrm{eff}, G}^{\rm (resum)} \ \equiv \ \frac{3 \, (m^2_G + \Pi_g)^2}{4 \, (16 \pi^2)}  \bigg[ \ln\bigg(\frac{m^2_G + \Pi_g}{\mu^2} \bigg) - \frac{3}{2}\bigg] \ - \ V_{\mathrm{eff}, G}^{\rm (d.c.)}\;, 
\end{equation}
where $V_{\mathrm{eff}, G}^{\rm (d.c.)}$ is the contribution of the double-counted diagrams.

\begin{figure}[t]
\centering
\includegraphics[width=0.9\textwidth]{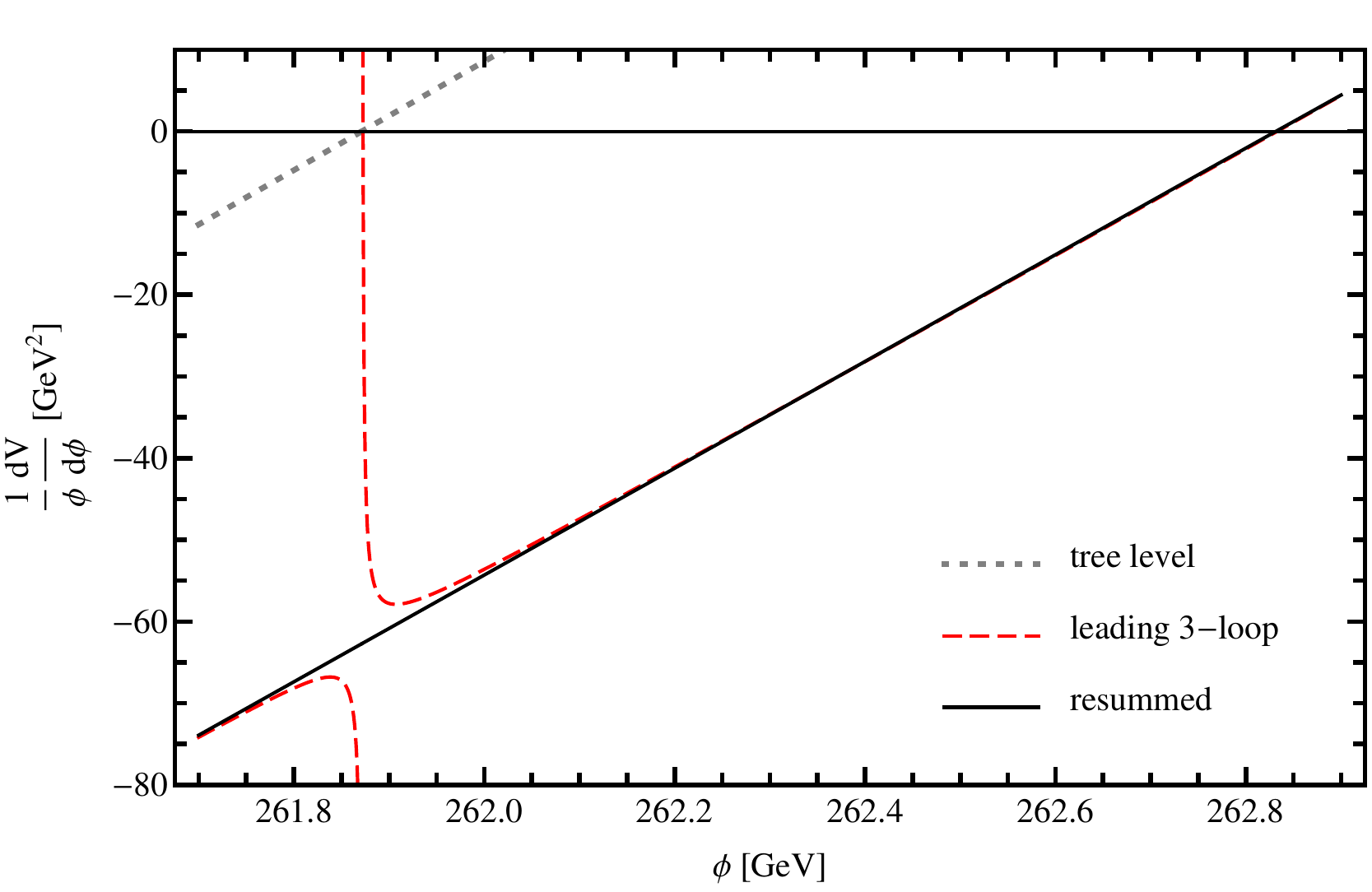}
\caption{Numerical estimates of $(1/\phi)\, d V_{\rm eff}/ d \phi$ versus~$\phi$, as obtained in different approaches for the scalar sector of the SM, using the $\overline{\rm MS}$ parameters: $\mu = 173.35\,\mathrm{GeV}$, $m = 93.36 \,\mathrm{GeV}$ and $\lambda = 0.12710$.
The dashed (red) line is the perturbative three-loop leading contribution, which exhibits an IR Goldstone-boson mass singularity at $\phi \approx 261.87$~GeV. The solid (black) line is the prediction of the approximate partial re\-summation prescription of~\cite{Martin:2014bca, Elias-Miro:2014pca}, whilst the dotted (gray) line corresponds to the tree-level contribution.\label{fig:IR_pert}}
\end{figure}

In~Figure~\ref{fig:IR_pert}, we present numerical estimates of the $\phi$-derivative of the 1PI effective potential, $d V_{\rm eff}/ d \phi$, as a function of~$\phi$, which are computed in two different approaches: (i)~the perturbative three-loop leading computation of ring diagrams consisting of two Goldstone-boson self-energy graphs as indicated in Figure~\ref{fig:IR} and (ii)~the approximate partial resum\-mation prescription followed in~\cite{Martin:2014bca, Elias-Miro:2014pca}. For definiteness, we use the following values for the $\overline{\rm MS}$ parameters: $\mu = 173.35\,\mathrm{GeV}$, $m = 93.36 \,\mathrm{GeV}$ and $\lambda = 0.12710$. As shown in~Figure~\ref{fig:IR_pert}, the dashed (red) line represents the perturbative three-loop leading contribution, which exhibits an IR Goldstone-boson mass singularity at $\phi \approx 261.87$~GeV, whereas the solid (black) line results from the approximate partial resummation prescription, which is ostensibly IR finite.  Finally, the dotted (gray) line corresponds to the tree-level result. In the next sections, we will develop a SI2PI approach to address the IR problem and compare the results of our approach with those using the approximate resummation prescription presented here.

\section{The SM Scalar Sector  in the 2PI Formalism}\label{sec:scalar}

In this section we briefly review the 2PI formalism applied to the scalar sector of the~SM, in the gaugeless limit of the theory. A pertinent discussion and further details may be found in \cite{Pilaftsis:2013xna, Pilaftsis:2015cka}. We postpone to Section~\ref{sec:ferm} the inclusion of chiral fermion quantum effects arising from top-quark Yukawa interactions.

Our starting point is the Lagrangian describing the scalar sector of the SM,
\begin{equation}\label{eq:lagr_scal}
\mathcal{L}_{\rm scalar} \ = \ (D^\mu \Phi^\dagger)(D_\mu \Phi) \; + \; m^2 \, \Phi^\dag \Phi \; - \; \lambda \, (\Phi^\dag \Phi)^2\;.
\end{equation}
In the above, $D_\mu = \partial_\mu + ig_w T^a W^a_\mu + ig' YB_\mu$ is the $SU(2)_L \times U(1)_Y$ covariant derivative, where $T^a = \sigma^a/2$ (with $a=1,2,3$) and $Y$ are the generators
of the $SU(2)_L$ and $U(1)_Y$ groups associated with the $W^a_\mu$ and $B_\mu$ gauge fields, respectively. In addition, $\Phi$ is the Higgs doublet with $Y = 1/2$, which is expanded about
the background field $\phi$ as
\begin{equation}
\Phi \ = \ \begin{pmatrix} G^+ \\
\frac{1}{\sqrt{2}}(\phi \,+\, H \,+\, i\, G^0)\end{pmatrix}\;,
\end{equation}
where $H$ is the observed Higgs boson, and $G^0$ and $G^+$ are the neutral and charged Goldstone bosons, respectively. In the $U(1)_Y$ gaugeless limit $g' \to 0$, the Lagrangian~\eqref{eq:lagr_scal} possesses a higher symmetry, the so-called custodial symmetry~\cite{Weinstein:1973gj, Weinberg:1979bn, Susskind:1978ms, Sikivie:1980hm}: $SU(2)_L \times SU(2)_R/Z_2 \simeq SO(4)$, which is spontaneously broken to a diagonal custodial subgroup $SU(2)_C\equiv SU(2)_{(L+R)}$.  As a consequence of the $SU(2)_C$ symmetry, the Goldstone bosons $G^0$ and $G^+$ are mass degenerate and their respective dressed propagators $\Delta^G$ and $\Delta^+$ are equal to each other.

The 2PI  effective action is obtained by introducing a local source $J(x)$, as in the usual 1PI effective action, and a bi-local source $K(x,y)$, with implicit $SU(2)$ group structure. By Legendre-transforming the connected generating functional with respect to these sources, one obtains the 2PI effective action $\Gamma[\phi,\Delta]$, depending on the background field $\phi$ and the \emph{dressed} propagators $\Delta$. In the gaugeless limit of the theory in which $g_w,g'\to 0$, the 2PI effective action for the SM scalar sector (expanded diagrammatically up to two-loop topology graphs) may conveniently be expressed as~\cite{Pilaftsis:2013xna, Pilaftsis:2015cka}
\begin{align}
\label{eq:gamma_scal}
\Gamma_{\rm scalar}^{(2)}[\phi,\Delta^H,\Delta^G,\Delta^+] \ &= \ \int \! \bigg[\frac{Z_0}{2} (\partial_\mu \phi)^2 \:+\:\frac{m^2 +
  \delta m_0^2}{2} \,\phi^2 \: -\: 
\frac{\lambda + \delta \lambda_0}{4}  \,\phi^4 \bigg] \notag \\[3pt]
& -\: \frac{i}{2} \Tr \big(\ln\Delta^H\big) \: -\: 
\frac{i}{2} \Tr \big(\ln \Delta^G \big) \: -\: 
i \Tr \big(\ln \Delta^+ \big) \displaybreak[0]\notag\\[3pt]
&- \: \frac{i}{2} \Tr \Big\{\Big[ Z_1 \, \partial^2\: +\: 
\big( 3 \lambda + \delta \lambda_1^A + 2 \delta \lambda_1^B\big)\,\phi^2
- \big( m^2 + \delta m^2_1\big) \Big]\,\Delta^H \Big\}  \notag\\[3pt]
&- \: \frac{i}{2} \Tr \Big\{\Big[ Z_1 \, \partial^2\: +\:
  \big(\lambda + \delta \lambda_1^A\big) \, \phi^2 - \big(m^2 + \delta m^2_1\big)
  \Big]\,\Delta^G \Big\}  \notag\\[3pt]
  &- \:\, i \, \Tr \Big\{\Big[ Z_1 \, \partial^2\: +\:
  \big(\lambda + \delta \lambda_1^A\big) \, \phi^2 - \big(m^2 + \delta m^2_1\big)
  \Big]\,\Delta^+ \Big\}\displaybreak[0]  \notag\\[3pt]
  & - \, \frac{i}{4} \, \bigg\{ -i(3 \lambda + \delta \lambda_2^A + 2 \delta \lambda_2^B) \int \! i\Delta^H i\Delta^H  \;-\; 2i (\lambda + \delta \lambda_2^A ) \int \! i\Delta^H i\Delta^G\notag\\ 
  & -4 i(\lambda + \delta \lambda_2^A ) \int \! i\Delta^H i\Delta^+  \;-\; i (3\lambda + \delta \lambda_2^A + 2 \delta \lambda_2^B) \int \! i\Delta^G i\Delta^G\notag\\ 
  & -4 i(\lambda + \delta \lambda_2^A ) \int \! i\Delta^G i\Delta^+  \;-\; 4 i (2\lambda + \delta \lambda_2^A + \delta \lambda_2^B) \int \! i\Delta^+ i\Delta^+ \bigg\}\displaybreak[0]\notag\\[-2mm]
  & - \, i \, \Bigg\{ \; \parbox{1.7cm}{\includegraphics[width=1.7cm]{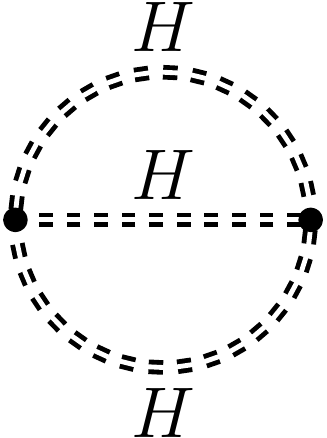}} \; + \; \parbox{1.7cm}{\includegraphics[width=1.7cm]{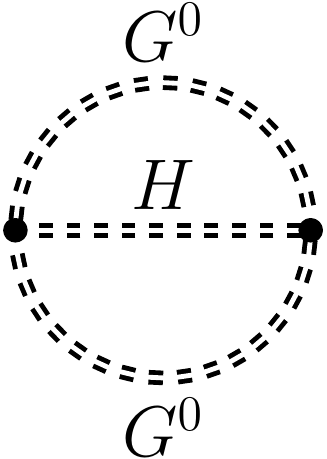}} \; + \; \parbox{1.7cm}{\includegraphics[width=1.7cm]{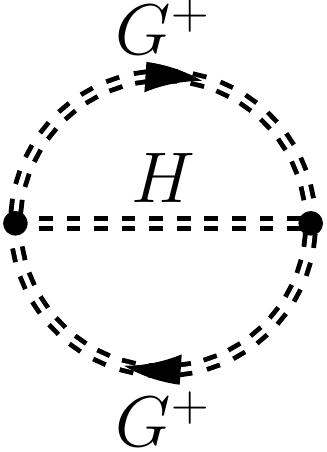}} \; \Bigg\} \;,
\end{align}
where the integrals are meant to be evaluated in position space over the common spacetime variable of the relevant fields and Green functions. In the last line of~\eqref{eq:gamma_scal}, double lines denote the dressed Green functions $\Delta$, whereas single lines are reserved to represent tree-level propagators.

The additional counter-terms (CTs) in~\eqref{eq:gamma_scal} that are not present in the 1PI formalism originate from the appearance of several related operators with mass-dimensions 2 and 4 in the 2PI effective action~\cite{vanHees:2001ik,Blaizot:2003an,Berges:2005hc}. Instead, the standard perturbative 1PI CTs appear at higher loop orders. In particular, no vertex CT is needed to be considered in the sunset diagrams in the last line of~\eqref{eq:gamma_scal}, as such a CT would be necessary to cancel subdivergences of higher-order diagrams. 

The Equation of Motions (EoMs) for the dressed propagators~$\Delta^{H,\,G,\,+}(k)$ are obtained by differentiating the 2PI effective action~$\Gamma_{\rm scalar}^{(2)}[\phi,\Delta^H,\Delta^G,\Delta^+]$ in~\eqref{eq:gamma_scal} with respect to these propagators. Because of the custodial $SU(2)_C$ symmetry of the SM scalar-sector 2PI effective action~$\Gamma_{\rm scalar}^{(2)}$, we can limit ourselves to the EoMs for $\Delta^H(k)$ and $\Delta^G(k)$, since $\Delta^+(k) = \Delta^G(k)$. After a Wick rotation to Euclidean space with Euclidean momentum~$k$, we obtain 
\begin{subequations}
   \label{eq:eomsbarescal} 
\begin{align}
  \Delta^{-1,\,H}(k) \ &= \ (1 + \delta Z_1) \, k^2 \:+\: (3 \lambda + \delta \lambda_1^A +
                         2\delta\lambda_1^B)\, \phi^2 \:-\: (m^2 + \delta m_1^2)  \notag\\[3pt]  &+\: (3
                                                                                                   \lambda + \delta\lambda_2^A + 2 \delta \lambda_2^B) \, \bm{\mathcal{T}}_{\!\!H} \:+\: 3 (\lambda + \delta \lambda_2^A) \,\bm{\mathcal{T}}_{\!\!G} \: - \:
                                                                                                   18 \lambda^2 \phi^2 \, \bm{\mathcal{I}}_{HH}(k) \:-\: 6 \lambda^2 \phi^2 \, \bm{\mathcal{I}}_{GG}(k) \;, 
  \label{eq:eombareHscal}
                                                                                                   \displaybreak[0]\\[9pt]
  \Delta^{-1,\,G}(k) \ &= \ (1 + \delta Z_1) \, k^2 \:+\: (\lambda + \delta \lambda_1^A )\, \phi^2 \:-\: (m^2 + \delta m_1^2) \notag\\[3pt]  &+\: (\lambda + \delta \lambda_2^A) \, \bm{\mathcal{T}}_{\!\!H} 
 \:+\:  (5
\lambda + 3 \delta\lambda_2^A + 2 \delta \lambda_2^B)\,\bm{\mathcal{T}}_{\!\!G} \: - \:
4 \lambda^2 \phi^2 \, \bm{\mathcal{I}}_{HG}(k) \;, \label{eq:eombareG0scal}
\end{align}
\end{subequations}
where $\delta Z_1 = Z_1 -1$. In writing down the two EoMs in~\eqref{eq:eomsbarescal}, we have introduced the tadpole and sunset integrals:
\begin{equation}\label{eq:TaIab}
\bm{\mathcal{T}}_{\!\!a} \ =\ \overline{\mu}^{2\epsilon} \int_p i \Delta^a(p)
\;, \qquad\qquad  
\bm{\mathcal{I}}_{ab}(k) \ =\ \overline{\mu}^{2\epsilon} \int_p
i \Delta^a(k + p) \, i \Delta^b(p) \;, 
\end{equation}
where $\ln\overline{\mu}^2 = \ln \mu^2 + \gamma - \ln(4\pi)$, with $\mu$ being the $\overline{\rm MS}$ renormalization mass scale. Here and in the following, the Latin indices run over $H,G,+$.

At the two-loop level of the 2PI effective action~$\Gamma_{\rm scalar}^{(2)}$, there is no wavefunction re\-normalization and so the CT $\delta Z_1$ can be set equal to zero. Otherwise, we may isolate the ultra-violet~(UV) divergences from the integrals~\eqref{eq:TaIab} by introducing the auxiliary propagator 
\begin{equation} 
  \label{eq:D0}
\Delta_0(k) \ \equiv \ (k^2 \; + \; \mu^2)^{-1}\;,
 \end{equation} 
which has the same asymptotic behaviour as the dressed propagators $\Delta^a(k)$. 
Given that $\Delta^a = \Delta_0 + O(\Delta_0^2)$, one may extract, for instance, the UV divergence of $\bm{\mathcal{I}}_{ab}(k)$ as 
\begin{equation} 
\bm{\mathcal{I}}_{ab}(k) \ = \ \overline{\mu}^{2\epsilon} \int_p \big[i \Delta_0(p)\big]^2 \ + \ \mathcal{I}_{ab}(k) \,, 
\end{equation}
 where $\mathcal{I}_{ab}(k)$ is the finite renormalized sunset integral. A more detailed discussion, including chiral fermion quantum effects, will be given in Section~\ref{sec:ferm} and~\ref{app:ren}.  The~EoMs~\eqref{eq:eomsbarescal} are renormalized by cancelling separately the subdivergences proportional to the re\-normalized tadpole integrals and the overall divergences proportional to the field powers~$\phi^0$ and~$\phi^2$~\cite{Fejos:2007ec,Pilaftsis:2013xna}. Out of $2 \times 4$ relations, only 5 of them are found to be independent, which uniquely fixes the value of the 5 CTs appearing in~\eqref{eq:eomsbarescal}. Hence, the renormalized EoMs are found to be~\cite{Pilaftsis:2013xna,Pilaftsis:2015cka} 
\begin{subequations}
   \label{eq:eomsren_scal} 
\begin{align}
  \Delta^{-1,\,H}(k) \ &= \ k^2 \:+\: 3 \lambda \phi^2 \,-\, m^2  \:+\: 3
                         \lambda  \, \mathcal{T}_H \:+\: 3 \lambda \,\mathcal{T}_G \: - \:
 18\lambda^2\phi^2 \, \mathcal{I}_{HH}(k) \notag\\
 &- \: 6\lambda^2\phi^2 \, \mathcal{I}_{GG}(k) \;+\; \Pi^{\mathrm{2PI}, (2)}_H \;, \\[6pt]
\Delta^{-1,\,G}(k) \ &= \ k^2 \:+\: \lambda \phi^2 \,-\, m^2  \:+\: \lambda  \, \mathcal{T}_H \:+\: 5 \lambda \,\mathcal{T}_G \: - \:
 4\lambda^2\phi^2 \, \mathcal{I}_{HG}(k) \:+ \: \Pi^{\mathrm{2PI}, (2)}_G  \;, 
\end{align}
\end{subequations}
where the analytic expression for the renormalized tadpole integral $\mathcal{T}_a$ is given in \cite{Pilaftsis:2015cka}. The same expression can also be inferred from~\eqref{eq:tad_ren}, for vanishing Yukawa couplings $h_t$. In~\eqref{eq:eomsren_scal} we have included also the renormalized two-loop 2PI self-energies $\Pi^{\mathrm{2PI}, (2)}_a$. However, as we discuss in more detail in Section~\ref{sec:ferm}, the latter contributions result from a three-loop order truncation of the 2PI effective action, and so we approximate them by their usual 1PI form evaluated in the zero-momentum limit $k\to 0$. Their analytic expressions are given in~\ref{app:int}.

We conclude this section by reminding the reader of an important feature of the SI2PI formalism \cite{Pilaftsis:2013xna} adopted here. The EoM for the background field $\phi$ is replaced by the standard 1PI WI stated later in~\eqref{eq:WI_phi}. In Section~\ref{sec:IR_2PI} we employ this WI to compute the SI2PI effective potential in terms of the dressed $G^0$ propagator.

\section{Quantum Effects from Chiral Fermions}\label{sec:ferm}

Our goal is now to include quantum effects from chiral fermions, by considering a simplified semi-perturbative 2PI framework with one third generation quark doublet and one right-handed top quark. This is  a non-trivial task within the SI2PI formalism. The inclusion of SM chiral fermions breaks down  explicitly the remaining custodial $SU(2)_C$ symmetry of the theory, giving rise to spurious custodially breaking effects that may even violate the Goldstone symmetry underpinning the SI2PI formalism. As we will see, however, such spurious effects can be consistently removed by appropriate renormalization, thereby reinforcing the Goldstone symmetry of the theory.

To start with, let us consider a simple but realistic extension of the SM scalar sector with a single
Yukawa coupling $h_t$ governing the interaction of the Higgs doublet $\Phi$ to third generation quarks. To be specific, the Yukawa interaction of interest is described by the Lagrangian
\begin{equation}
-\, \mathcal{L}_{\rm Y} \ = \ h_t \, \varepsilon^{ab} \,\overline{Q}_{L,a} \Phi_b^\dagger \, t_R \; +\; \mathrm{H.c}. \,,
\end{equation}
where $\varepsilon^{ab}$ is the antisymmetric Levi-Civita symbol, $Q_L = (t_L \,\,b_L)^{\mathtt{T}}$ is the left-handed $SU(2)_L$ quark doublet of the third generation and $t_R$ the right-handed top quark.

For the purpose of this study, we consider a semi-perturbative approach to include chiral fermion quantum effects in the 2PI effective action. Specifically, we only couple the scalar fields to bilocal sources, but not the chiral fermions. Upon a Legendre transform with respect to these bilocal sources, we generate dressed Green functions for all SM scalar fields, but not for the chiral fermions, i.e.~for the third generation quarks. In this simplified framework, the 2PI effective action expanded to two-loop order is given by
\begin{align}
  \label{eq:Gamma}
\Gamma^{(2)}[\phi,\Delta^H,\Delta^G,\Delta^{+}] \ &= \ \Gamma_{\mathrm{scalar}}^{(2)}[\phi,\Delta^H,\Delta^G,\Delta^{+}] \; + \; 3\, i \, \Tr \ln S^{\alpha\,(0)}[\phi] \notag\\[3pt]
& - \, i \, \Bigg\{ \; \parbox{1.7cm}{\includegraphics[width=1.7cm]{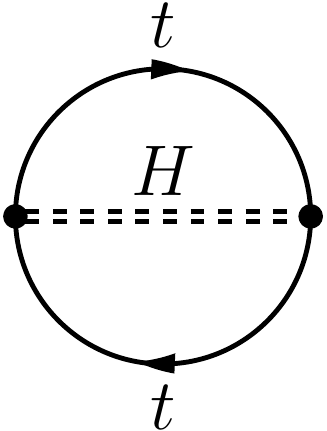}} \; + \; \parbox{1.7cm}{\includegraphics[width=1.7cm]{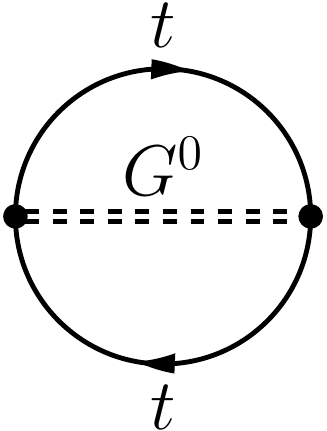}} \; + \; \parbox{1.7cm}{\includegraphics[width=1.7cm]{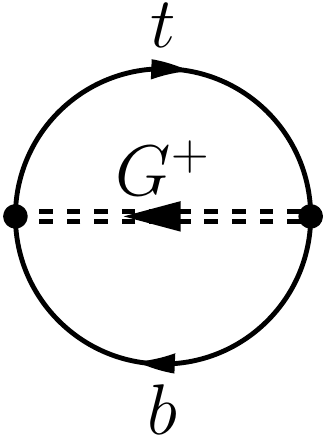}} \; \Bigg\} \;,
\end{align}
where $\Gamma_{\rm scalar}^{(2)}$ is the 2PI effective action for the SM scalar sector given by~\eqref{eq:gamma_scal} and $S^{\alpha \, (0) }[\phi]$, with $\alpha=t,b$, is the tree-level fermion propagator. The factor of 3 in~\eqref{eq:Gamma} arises from the sum over degenerate colour degrees of freedom. The 2PI effective action~$\Gamma^{(2)}$ can be renormalized in a fashion similar to the scalar case discussed in Section~\ref{sec:scalar}, by introducing a set of renormalized parameters and their associated CTs. Notice that the wavefunction renormalization $Z_1$ in $\Gamma_{\mathrm{scalar}}^{(2)}$ can no longer be set to 1, because it is needed to renormalize the UV divergences of the diagrams involving fermions.

\vfill\eject

Differentiating $\Gamma^{(2)}[\phi,\Delta^H,\Delta^G,\Delta^{+}]$ in~\eqref{eq:Gamma}  with respect to 
$\Delta^H(k)$, $\Delta^G(k)$ and $\Delta^{+}(k)$, we obtain respectively the EoMs (expressed in the Euclidean momentum space)
\begin{subequations}
   \label{eq:eomsbare}
\begin{align}
\Delta^{-1,\,H}(k) \ &= \ (1 + \delta Z_1) \, k^2 \:+\: (3 \lambda + \delta \lambda_1^A +
2\delta\lambda_1^B)\, \phi^2 \:-\: (m^2 + \delta m_1^2)  \notag\\[3pt]  &+\: (3
\lambda + \delta\lambda_2^A + 2 \delta \lambda_2^B) \, \bm{\mathcal{T}}_{\!\!H} \:+\: (\lambda + \delta \lambda_2^A) \,\bm{\mathcal{T}}_{\!\!G} \:+\: 2 \,(\lambda + \delta \lambda_2^A) \,\bm{\mathcal{T}}_{\!\!+}\: - \:
18 \lambda^2 \phi^2 \, \bm{\mathcal{I}}_{HH}(k) \notag\\[3pt] 
&-\: 2 \lambda^2 \phi^2 \, \bm{\mathcal{I}}_{GG}(k) 
\:-\: 4 \lambda^2 \phi^2 \, \bm{\mathcal{I}}_{++}(k) \:+\: \bm{\Sigma}_H(k)\;, \label{eq:eombareH}
\displaybreak[0]\\[9pt]
\Delta^{-1,\,G}(k) \ &= \ (1 + \delta Z_1) \, k^2 \:+\: (\lambda + \delta \lambda_1^A )\, \phi^2 \:-\: (m^2 + \delta m_1^2)  \:+\: (\lambda + \delta \lambda_2^A) \, \bm{\mathcal{T}}_{\!\!H} \notag\\[3pt] 
&\quad \:+\:  (3
\lambda + \delta\lambda_2^A + 2 \delta \lambda_2^B)\,\bm{\mathcal{T}}_{\!\!G} \:+\: 2 \,(\lambda + \delta \lambda_2^A) \,\bm{\mathcal{T}}_{\!\!+}\: - \:
4 \lambda^2 \phi^2 \, \bm{\mathcal{I}}_{HG}(k) \:+\: \bm{\Sigma}_G(k)\;, \label{eq:eombareG0}
\displaybreak[0]\\
\Delta^{-1,\,+}(k) \ &= \ (1 + \delta Z_1) \, k^2 \:+\: (\lambda + \delta \lambda_1^A + \delta \lambda_1^{\rm cb})\, \phi^2 \:-\: (m^2 + \delta m_1^2)  \:+\: (\lambda + \delta \lambda_2^A) \, \bm{\mathcal{T}}_{\!\!H} \notag\\[3pt] 
&\quad \:+\:  (\lambda + \delta \lambda_2^A)\,\bm{\mathcal{T}}_{\!\!G} \:+\: 2 \,(2
\lambda + \delta\lambda_2^A + \delta \lambda_2^B) \,\bm{\mathcal{T}}_{\!\!+}\: - \:
4 \lambda^2 \phi^2 \, \bm{\mathcal{I}}_{H+}(k) \:+\: \bm{\Sigma}_+(k)\;. \label{eq:eombareGp}
\end{align}
\end{subequations}
In the above, $\bm{\Sigma}_a(k)$ (with $a = H,G,+$) are the one-loop $H$-, $G^0$- and $G^+$-boson self-energies, respectively, which have been calculated in the standard perturbative 1PI formalism.
The one-loop self-energies $\bm{\Sigma}_a(k)$ are renormalized in the $\overline{\rm MS}$ scheme. Upon $\overline{\rm MS}$ renormalization, their analytic expressions in the Euclidean momentum space are given by
\begin{subequations}\label{eq:sigmas}
\begin{align}
\Sigma_H(k) \ &= \ - \frac{3 \, h_t^2}{16 \pi^2} \Big( s\, B(k;t,t) \:-\: 4 \, t B(k;t,t) \:+\: 2 \,A(t) \Big) \;,\\
\Sigma_G(k) \ &= \ - \frac{3 \, h_t^2}{16 \pi^2} \Big( s\, B(k;t,t) \:+\: 2 \,A(t) \Big) \;,\\
\Sigma_+(k) \ &= \ - \frac{3 \, h_t^2}{16 \pi^2} \Big( s\, B(k;t,0) \:-\:  t B(k;t,0) \:+\: A(t) \Big) \;.
\end{align}
\end{subequations}
Here, we have introduced the $\phi$-dependent tree-level top-quark mass squared $t = h_t^2 \phi^2/2$ and the kinematic variable $s = -k^2 \leq 0$. In addition, by analytic continuation,
we allow $s$ to assume positive values as well, in which case $s$ may be identified with the usual time-like Mandelstam variable. Finally, the $\overline{\rm MS}$-renormalized one-loop functions $A(x)$ and $B(k;x,y)$ are defined in~\ref{app:int}.

\begin{figure}
\centering
$\parbox{3cm}{\includegraphics[height=4.53cm]{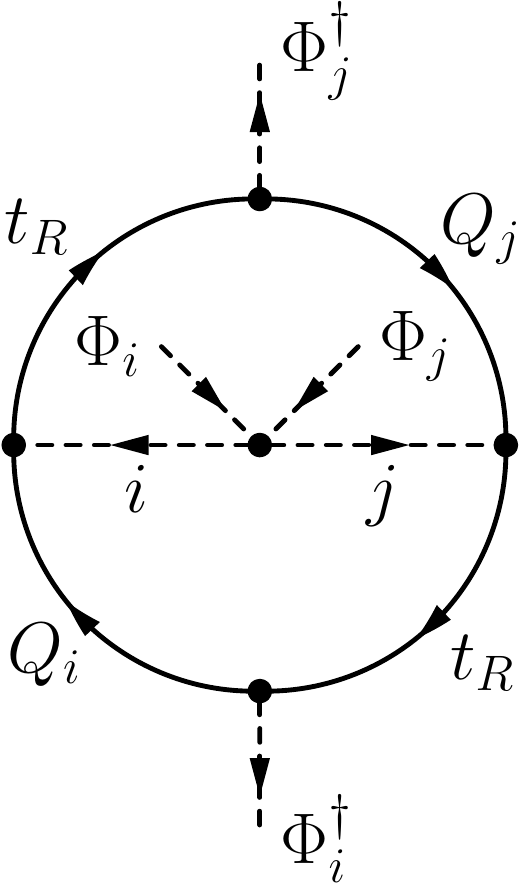}} \quad + \quad\;\;\; \parbox{3.1cm}{\includegraphics[height=4.53cm]{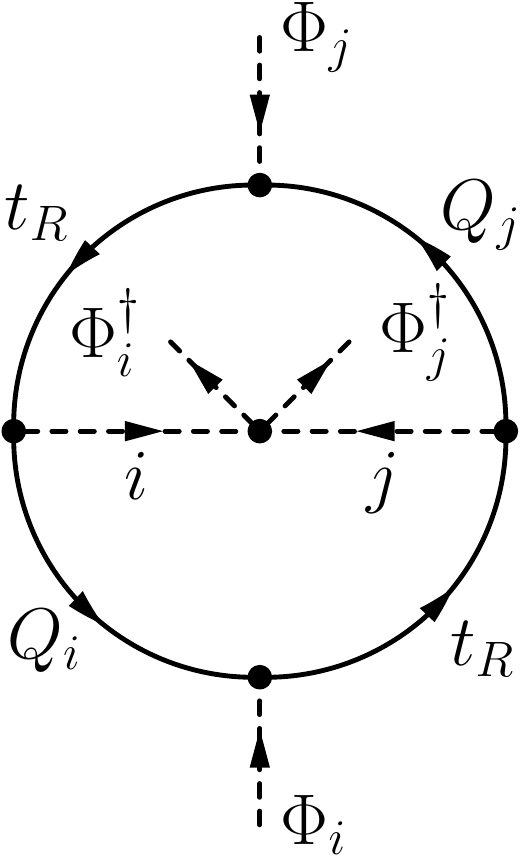}} \quad \sim \quad \lambda \, h_t^4  \, (\Phi^\dag \Phi)^2$
\caption{Leading two-loop topology for the operator $(\Phi^\dag \Phi)^2$ that gives rise to 
custodially violating artifacts in a fixed loop-order truncated 2PI effective action at zero momentum.  See text for details.\label{fig:2loop_topo}}
\end{figure}

\begin{figure}
\centering
$\parbox{3.9cm}{\includegraphics[height=5.1cm]{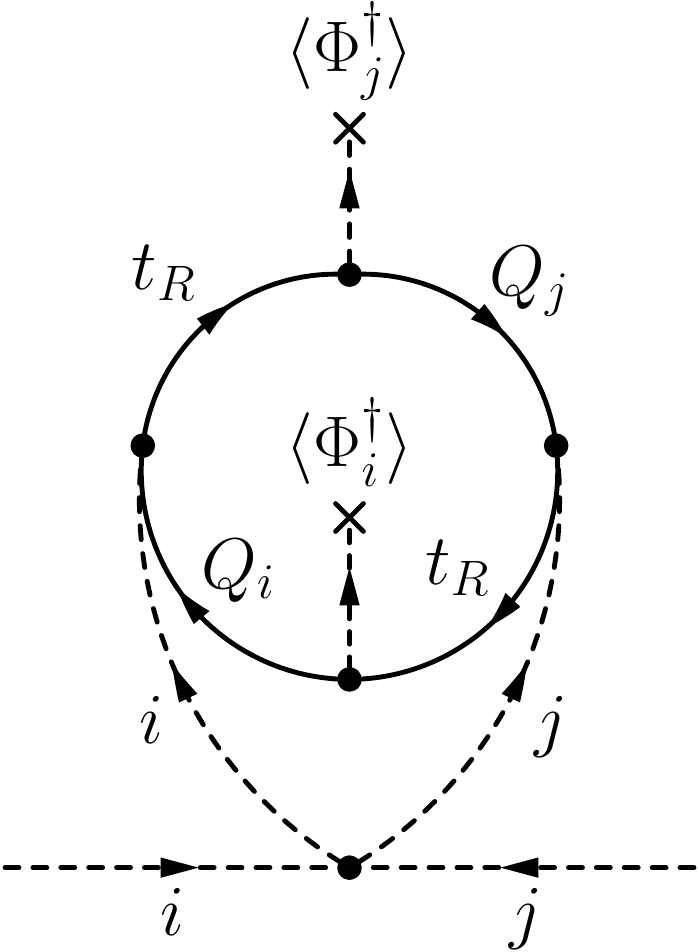}} \;+\; \parbox{3.9cm}{\includegraphics[height=5.1cm]{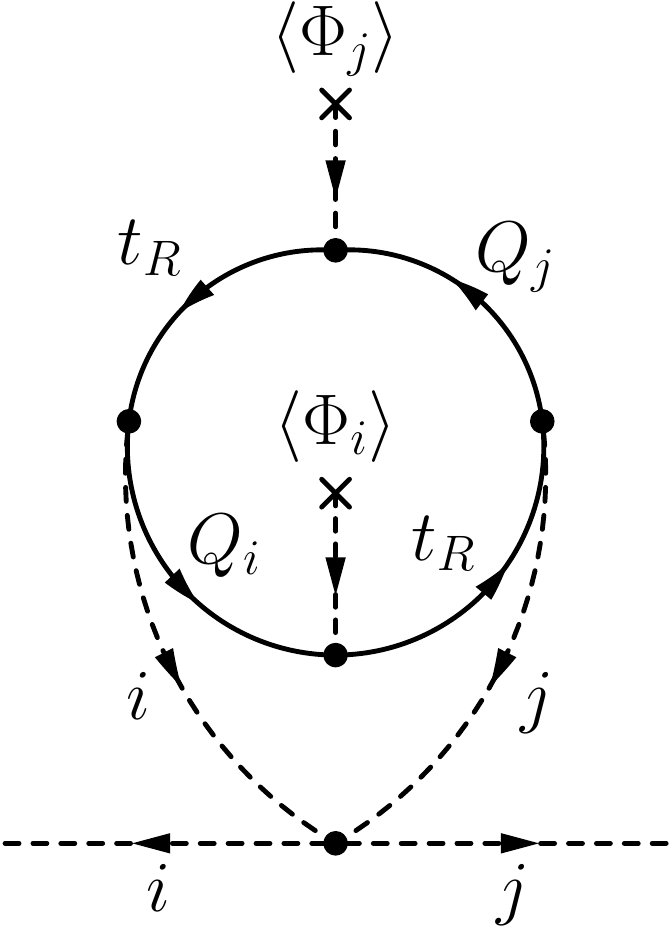}} \,
\begin{aligned}[t]
 & = \;C \, \Phi_i \Phi_j \langle \Phi^\dagger_i \rangle \langle \Phi_j^\dagger \rangle \;+\; C \, \Phi_i^\dag \Phi_j^\dag \langle \Phi_i \rangle \langle \Phi_j \rangle \\[6pt]
 & \supset \ - \, C \, \phi^2 \, G^0 G^0 
 \end{aligned}$ 
\caption{Field contractions of the topology in Figure~\ref{fig:2loop_topo} that generate Goldstone-boson self-energy graphs that are included in the one-loop 2PI resummation.\label{fig:2loop_incl}}
\end{figure}

In addition to the CTs that occur in the 2PI effective action for the SM scalar sector, one extra CT, $\delta \lambda_1^{\rm cb}$, needs to be considered when chiral fermion quantum effects are included in the EoM~\eqref{eq:eombareGp}.  As mentioned above, top Yukawa interactions break explicitly the custodial $SU(2)_C$ symmetry, leading to artifacts that violate the Goldstone symmetry in the 2PI effective action and so the equality between the dressed Goldstone-boson propagators, i.e.~$\Delta^G (0) = \Delta^+ (0)$, at zero momentum $k=0$.  We should stress here that these artifacts arise from
a fixed loop-order truncation of the 2PI effective action and are absent in the standard 1PI perturbative formulation of QFT. In particular, one can show that some of the UV divergences proportional to $\phi^2$ get missed in the EoMs, because of this finite loop-order truncation of the 2PI action. 

\begin{figure}
\centering
\begin{align*}
&\parbox{4.6cm}{\includegraphics[width=4.53cm]{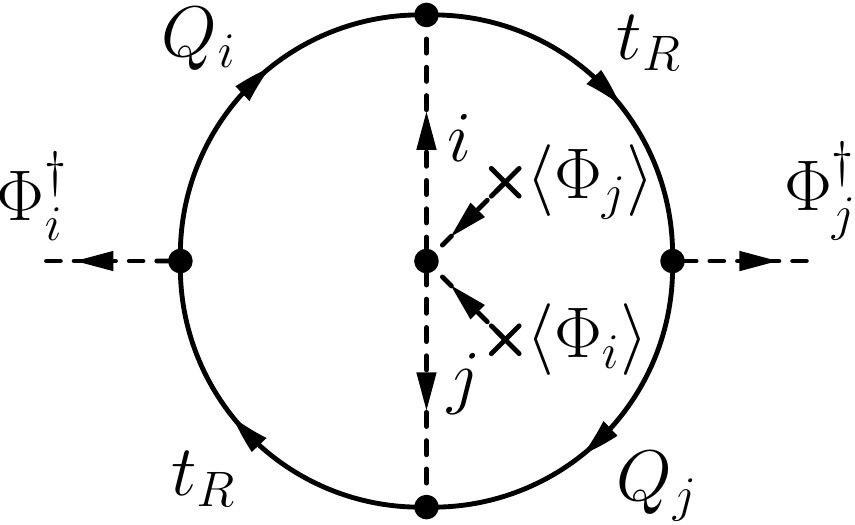}} \;+\; \parbox{4.6cm}{\includegraphics[width=4.53cm]{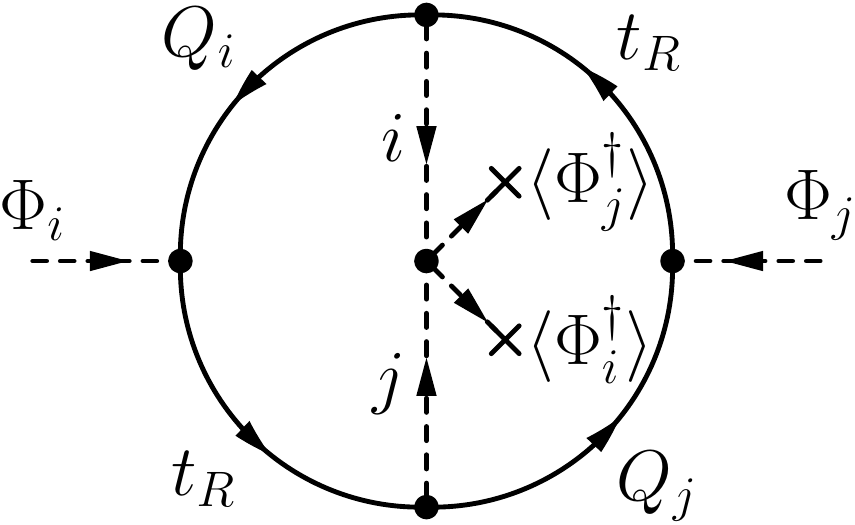}} \,
\begin{aligned}[t]
 & = \;C \, \langle \Phi_i \rangle \langle \Phi_j \rangle \Phi^\dagger_i  \Phi_j^\dagger  \\[6pt]
 &+\; C \, \langle \Phi_i^\dag \rangle \langle \Phi_j^\dag \rangle  \Phi_i  \Phi_j   \ \supset \ - \, C \, \phi^2 \, G^0 G^0 
 \end{aligned}\\[1cm]
&\parbox{4.6cm}{\includegraphics[width=4.53cm]{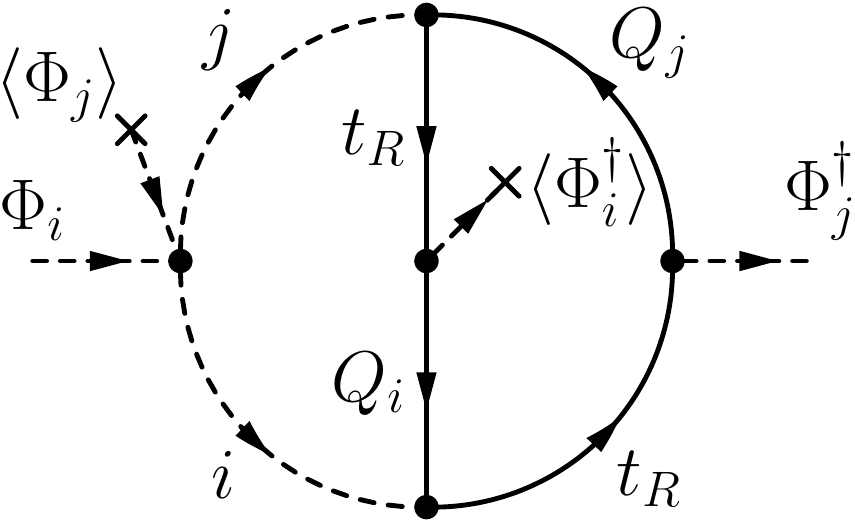}} \;+\; \parbox{4.6cm}{\includegraphics[width=4.53cm]{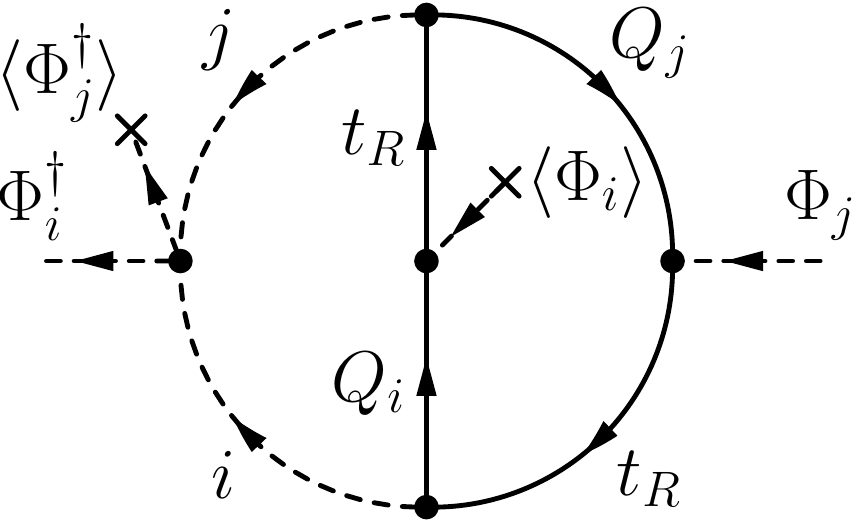}} \quad + \quad (i \leftrightarrow j)\\[9pt]
& \hspace{11em} = \; C \, \Phi_i \langle \Phi_j \rangle \langle \Phi^\dagger_i \rangle  \Phi_j^\dagger \;+\; C \, \Phi_i^\dag \langle \Phi_j^\dag \rangle  \langle \Phi_i \rangle \Phi_j  \; + \; (i \leftrightarrow j) \ \supset \ 2\,C \, \phi^2 \, G^0 G^0
\end{align*}
\caption{Field contractions of the topology in Figure~\ref{fig:2loop_topo} that generate Goldstone-boson self-energy graphs that are {\em not} included in the one-loop 2PI resummation. In perturbation theory, these contributions cancel against the ones in Figure~\ref{fig:2loop_incl} at zero  momentum.\label{fig:2loop_notincl}}
\end{figure}

To understand better the mechanism of {\em hard} custodial violation in the 2PI action at zero momentum, let us consider the two-loop topology in Figure~\ref{fig:2loop_topo}. This topology generates an operator of the form $\Phi^\dagger_i \Phi^\dagger_j \Phi_i \Phi_j$. When two of the external fields are taken as the background field $\phi$, the field contractions involving one index $i$ and one index $j$ would give rise to a hard custodial violation, i.e.~to different contributions for the neutral and the charged Goldstone-boson self-energies at zero momentum. In perturbation theory these contributions vanish, because all field contractions are included at a fixed given loop order. However, in a two-loop truncated 2PI effective action, the resulting two-loop self-energies in Figure~\ref{fig:2loop_incl} are included because of the 2PI resummation, but not the ones in Figure~\ref{fig:2loop_notincl}. As shown in Figures~\ref{fig:2loop_incl} and~\ref{fig:2loop_notincl}, all the different field contractions are needed to make the spurious custodially-violating terms vanish at zero momentum.  In particular, some of the UV divergences proportional to~$\phi^2$ get  missed in~\eqref{eq:eombareG0} and~\eqref{eq:eombareGp}, thus spoiling the equality $\Delta^G (0) = \Delta^+ (0)$, at zero momentum $k=0$. Hence, the inclusion of $\delta \lambda_1^{\rm cb}$ is necessary to compensate for this artifact. Moreover, the finite part of $\delta \lambda_1^{\rm cb}$ can be chosen, such that the renormalized neutral and charged Goldstone-boson propagators have both massless poles at $\phi=v$, thereby reinforcing the Goldstone symmetry of the theory within the context of the SI2PI formalism adopted in this paper.

To renormalize the EoMs stated in~\eqref{eq:eomsbare}, we deploy the same strategy as for the scalar case presented in the previous section, which closely follows~\cite{Pilaftsis:2013xna}. The only new aspect is that fermion quantum loops as described by the self-energies $\Sigma_a(k)$ modify the asymptotic behaviour of the dressed scalar propagators $\Delta^{H,\,G,\,+}(k)$. In the following, we show how the scalar sunset integrals~$\bm{\mathcal{I}}_{ab}$ get renormalized. The renormalization of the tadpole integrals~$\bm{\mathcal{T}}_{\!\!a}$ goes along similar lines, but it is technically more involved and will therefore be discussed in detail in~\ref{app:ren}.

In order to isolate the UV divergences appearing in the loop integrals involving scalar dressed propagators, we introduce the auxiliary propagator 
\begin{equation}
   \label{eq:aux} 
\widetilde{\Delta}_0(k) \ \equiv \ \left(k^2 \;+\; \mu^2 \;+\; \widetilde{\Sigma}(k)\,\right)^{-1}\;, \end{equation} 
where $\widetilde{\Sigma}(k)$ is chosen so as to have the same functional form as the neutral Goldstone-boson self-energy involving a top-quark loop, but with a fictitious $\overline{\rm MS}$ mass $\mu$, i.e.  
\begin{equation} 
\widetilde{\Sigma}(k) \ = \ - \frac{3 \, h_t^2}{16 \pi^2} \Big( s\, B(k;\mu^2,\mu^2) \:+\: 2 \,A(\mu^2) \Big) \;.  
\end{equation} 
Notice that all the self-energies~\eqref{eq:sigmas} have the same asymptotic behaviour as $\widetilde{\Sigma}(k)$, for high values of~$s$. As a consequence, the auxiliary propagator has the same asymptotic behaviour as that of the scalar dressed propagators, i.e.  
\begin{equation}
  \label{eq:expans1} 
\Delta^a(k) \ = \ \widetilde{\Delta}_0(k)\; + \; O\left(\widetilde{\Delta}_0^2(k)\right) \;.  
\end{equation} 
Our aim is to find a set of CTs in terms of the auxiliary propagator $\widetilde{\Delta}_0(k)$. Since the so-derived CTs will only depend on the $\overline{\rm MS}$ mass $\mu$ and the parameters of the Lagrangian, the EoMs will be successfully renormalized, for any value of the field $\phi$~\footnote{Observe that this procedure guarantees that the EoMs are successfully renormalized also at finite temperature $T$ with $T$-independent counterterms, as they should be.}. 

With the aid of the auxiliary propagator~$\widetilde{\Delta}_0(k)$, we may now extract the UV-divergent part of the sunset integral $\bm{\mathcal{I}}_{ab}(k)$ by introducing the loop integral
\begin{equation} 
  \widetilde{\bm{\mathcal{I}}}_0 \ \equiv \ \overline{\mu}^{2 \epsilon} 
  \int_k \; \widetilde{\Delta}^2_0(k) \;.  
\end{equation} 
We note that the combination $\bm{\mathcal{I}}_{ab}(k) - \widetilde{\bm{\mathcal{I}}}_{0}$ is finite, because of the expansion~\eqref{eq:expans1}. In order to exactly match our results to the perturbative $\overline{\rm MS}$-scheme results at two-loop accuracy, the finite part of the CTs has to be chosen accordingly. More explicitly, the CTs have to contain only the UV poles of $\widetilde{\bm{\mathcal{I}}}_0$ and not finite constant terms, when $\widetilde{\bm{\mathcal{I}}}_0$ is expanded perturbatively at two-loop order. To  achieve this, we subtract from $\widetilde{\bm{\mathcal{I}}}_0$ its finite piece in a two-loop $\overline{\rm MS}$-renormalization, which we denote as $\widetilde{{\mathcal{I}}}_0\big|^{(2)}_{\rm fin}$, as follows:
\begin{align}
                                                          \widetilde{{\mathcal{I}}}_0\big|^{(2)}_{\rm fin} \ & = \ \Bigg[ \; \parbox{1.7cm}{\includegraphics[width=1.7cm]{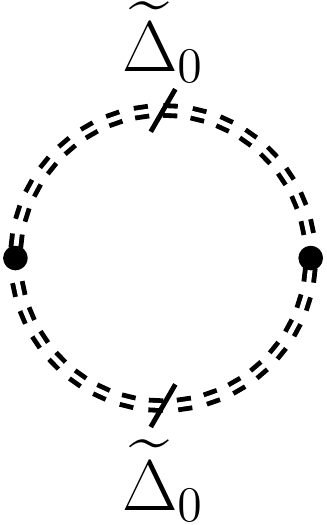}} \; \Bigg]^{(2)}_{\rm fin} \ = \ \Bigg[ \; \parbox{1.7cm}{\includegraphics[width=1.7cm]{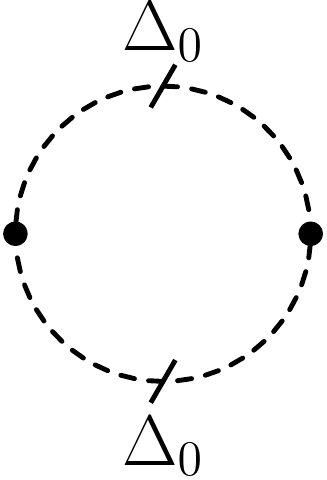}} \; + \; 2 \; \parbox{2.1cm}{\includegraphics[width=2.1cm]{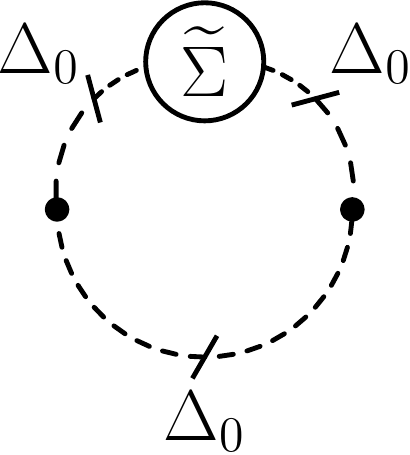}} \;\; \Bigg]^{(2)}_{\rm fin} \notag\\
                                                                                                             &= \ \Bigg[ \; 2 \; \parbox{2.1cm}{\includegraphics[width=2.1cm]{D0D0SigmatD0}} \;\; \Bigg]_{\rm fin} \;.
\end{align} 
In arriving at the last equality, we have used the fact that the perturbative one-loop diagram has no
finite terms within the $\overline{\rm MS}$ scheme, whilst the last finite two-loop diagram is calculated explicitly in~\ref{app:ren} [cf.~\eqref{eq:fin_int_b}]. Thus, in the $\overline{\rm MS}$ scheme, the UV infinite part of the sunset integral is extracted as 
\begin{equation}
  \label{eq:ICT} 
\widetilde{\bm{\mathcal{I}}}_{\rm CT} \ \equiv \ \widetilde{\bm{\mathcal{I}}}_0 \; - \; \widetilde{{\mathcal{I}}}_0\big|^{(2)}_{\rm fin} \;.
\end{equation} 
Hence, the unrenormalized sunset integral $\bm{\mathcal{I}}_{ab}$ may be written down as the sum
\begin{equation}
  \label{eq:sunset_part} 
  \bm{\mathcal{I}}_{ab} \ = \ \widetilde{\bm{\mathcal{I}}}_{\rm CT} \; + \; \mathcal{I}_{ab} \;,  
\end{equation} 
where $\mathcal{I}_{ab}$ is the corresponding $\overline{\rm MS}$-renormalized sunset integral given by
\begin{equation}
  \label{eq:ren_sun} 
\mathcal{I}_{ab}(k) \ = \ \int_p \bigg( \Delta^a(p-k)\, \Delta^b(p) \:-\: \widetilde{\Delta}^2_0(p)\bigg) \ + \ \widetilde{{\mathcal{I}}}_0\big|^{(2)}_{\rm fin} \;,
 \end{equation} 
with $\widetilde{{\mathcal{I}}}_0\big|^{(2)}_{\rm fin}$ given by~\eqref{eq:fin_int_b}. An analogous partitioning approach to the $\overline{\rm MS}$-renormalization of the tadpole integrals~$\mathcal{T}_a$ is presented in~\ref{app:ren}. 

Proceeding as in the scalar case, we may now impose the vanishing of all UV divergences on the EoMs~\eqref{eq:eomsbare}, which are contained in the CTs and the UV-infinite parts of the integrals as discussed above. We note that in our simplified semi-perturbative framework, the wavefunction renormalization $\delta Z_1$ of the dressed scalar propagators $\Delta^{H\,,G,\, +}(k)$ may be calculated, within the context of standard perturbation theory. Instead, wave\-function re\-normalizations for fermions do not enter the one-loop order EoMs for $\Delta^{H\,,G,\, +}(k)$ and therefore they do not need to be considered here.  Otherwise, exactly as done in the scalar case discussed in the previous section, one needs to require that the UV-divergent terms proportional to~$\phi^2$, $\mathcal{T}_a$ and the remaining overall divergences vanish. These conditions also ensure that all sub\-divergences
get cancelled. Out of $3 \times 5$ relations, only 6 of them are found to be independent, which are sufficient to fix the values of the 6 CTs~$\delta m^2_1$, $\delta \lambda_1^{A,B,{\rm cb}}$ and $\delta \lambda_2^{A,B}$ in terms of (UV divergent) integrals involving the auxiliary propagator $\widetilde{\Delta}_0(k)$, the $\overline{\rm MS}$ mass $\mu$ and Lagrangian parameters, such as $m^2$, $\lambda$ and $h_t$. Explicit analytic expressions for all these CTs are exhibited in~\ref{app:ren}.

After executing the above renormalization programme, the renormalized EoMs for the dressed propagators $\Delta^{H\,,G,\, +}(k)$ are found to be (in Euclidean $k$-momentum representation) \begin{subequations}
   \label{eq:eomsren} 
\begin{align}
  \Delta^{-1,\,H}(k) \ &= \ k^2 \:+\: 3 \lambda \phi^2 \,-\, m^2  \:+\: 3
                         \lambda  \, \mathcal{T}_H \:+\: \lambda \,\mathcal{T}_G \:+\: 2 \lambda \,\mathcal{T}_+ \: - \:
 18\lambda^2\phi^2 \, \mathcal{I}_{HH}(k) \notag\\
 &- \: 2\lambda^2\phi^2 \, \mathcal{I}_{GG}(k) \: - \:
 4\lambda^2\phi^2 \, \mathcal{I}_{++}(k) \: + \: \Sigma_H(k) \: + \: \Pi^{\mathrm{2PI}, (2)}_H \: + \: \Sigma^{\mathrm{2PI}, (2)}_H \;,\\[8pt]
\Delta^{-1,\,G}(k) \ &= \ k^2 \:+\: \lambda \phi^2 \,-\, m^2  \:+\: \lambda  \, \mathcal{T}_H \:+\: 3 \lambda \,\mathcal{T}_G \:+\: 2 \lambda \,\mathcal{T}_+ \: - \:
 4\lambda^2\phi^2 \, \mathcal{I}_{HG}(k) \notag\\
 &+ \: \Sigma_G(k) \: + \: \Pi^{\mathrm{2PI}, (2)}_G\: + \: \Sigma^{\mathrm{2PI}, (2)}_G  \;,\\[8pt]
\Delta^{-1,\,+}(k) \ &= \ k^2 \:+\: \lambda \phi^2 \,-\, m^2  \:+\: \lambda  \, \mathcal{T}_H \:+\: \lambda \,\mathcal{T}_G \:+\: 4 \lambda \,\mathcal{T}_+ \: - \:
 4\lambda^2\phi^2 \, \mathcal{I}_{H+}(k) \notag\\
 &+ \: \Sigma_+(k) \: + \: \Pi^{\mathrm{2PI}, (2)}_+\: + \: \Sigma^{\mathrm{2PI}, (2)}_+\;,
\end{align}
\end{subequations}
where the analytic expressions for $\Sigma_a(k)$, $\mathcal{I}_{ab}(k)$ and $\mathcal{T}_a$ are given in~\eqref{eq:sigmas}, \eqref{eq:ren_sun} and~\eqref{eq:tad_ren}, respectively. As done in the scalar case discussed in Section~\ref{sec:scalar}, we have also included the renormalized two-loop self-energies $\Pi^{\mathrm{2PI}, (2)}_a(k)$ and $\Sigma^{\mathrm{2PI}, (2)}_a(k)$, which are obtained from three-loop 2PI vacuum diagrams, upon cutting scalar propagator lines and approximating the resulting self-energies by their perturbative 1PI forms in the zero-momentum limit $k \to 0$~\footnote{Notice that the pure scalar two-loop corrections to the $G^0$- and $G^+$-boson self-energies are equal for any value of the momentum~$k$,  because of the custodial $SU(2)_C$ symmetry, i.e.~$\Pi^{\mathrm{2PI}, (2)}_G(k) = \Pi^{\mathrm{2PI}, (2)}_+(k)$.}. The expressions for the above two-loop 1PI self-energies are given in~\ref{app:int}. 
Within this approximated SI2PI framework, we are now able to consistently compare our results in the next section with those obtained in a full perturbative two-loop calculation of the SM effective potential in the gaugeless limit of the theory.

\section{Symmetry-Improved 2PI Approach to Resumming IR Divergences}\label{sec:IR_2PI}

In this section we consider the SI2PI formalism proposed in~\cite{Pilaftsis:2013xna} to study the problem of IR divergences in the SM effective potential. The SI2PI formalism is a rigorous and self-consistent theoretical framework, and proves suitable to address the Goldstone-boson~IR problem outlined in Section~\ref{sec:IR} for a number of reasons. First, it is a first-principle method for performing diagrammatic resummations, without the need to resort to {\it ad hoc} subtractions in order to achieve single counting of graphs. Second, as we will explicitly demonstrate below, the 2PI nature of our approach takes into account more topologies of graphs, as well as the momentum dependence of the self-energy insertions that are resummed.  In this respect, the SI2PI approach differs from the approximate resummation prescription of~\cite{Martin:2014bca, Elias-Miro:2014pca}.  Finally, as discussed in detail in~\cite{Pilaftsis:2013xna}, the dressed Higgs- and Goldstone-boson propagators exhibit the proper threshold properties within the SI2PI formalism, which originate from the kinematic opening of on-shell multi-particle states in the loops. In particular, the Goldstone bosons~$G^{0,+}$ in quantum loops are exactly massless at the radiatively corrected VEV of the background field~$\phi$, since they are mediated by the resummed propagators~$\Delta^{G,\,+}(k)$ which have massless poles. Nevertheless, it can be shown that the SI2PI effective potential of the gaugeless SM has no IR infinities.

\begin{figure}[t]
\centering
\begin{align*}
\Delta^{-1}(k;\,\phi) \quad &= \quad   {\Delta^{(0)\,-1}(k;\,\phi)} \quad + \quad \parbox{1.13cm}{\includegraphics[width=1.13cm]{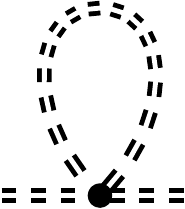}} \quad + \quad \parbox{1.7cm}{\includegraphics[width=1.7cm]{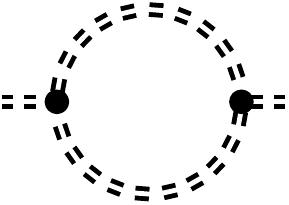}} \quad + \quad \parbox{1.7cm}{\includegraphics[width=1.7cm]{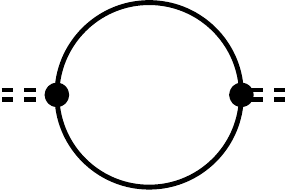}} \\[4mm]
 &\qquad + \; \Bigg[\;\parbox{1.7cm}{\includegraphics[width=1.7cm]{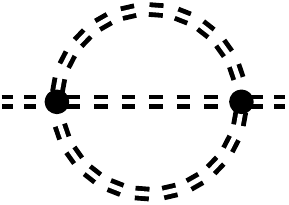}} \; + \; \parbox{1.7cm}{\includegraphics[width=1.7cm]{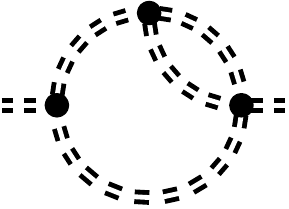}} \; + \; \parbox{1.7cm}{\includegraphics[width=1.7cm]{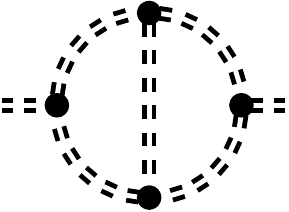}} \; + \; \parbox{2.27cm}{\includegraphics[width=2.27cm]{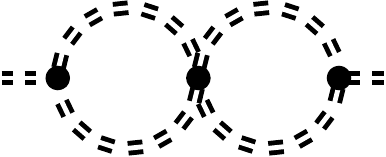}} \\[4mm]
 &\qquad + \quad\, \parbox{1.7cm}{\includegraphics[width=1.7cm]{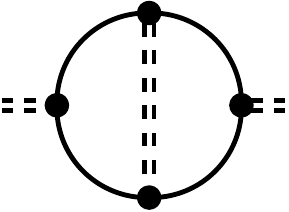}} \; + \;\parbox{1.7cm}{\includegraphics[width=1.7cm]{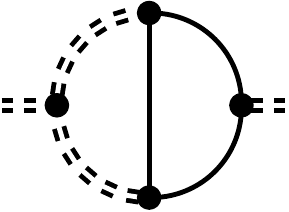}} \; + \; \parbox{1.7cm}{\includegraphics[width=1.7cm]{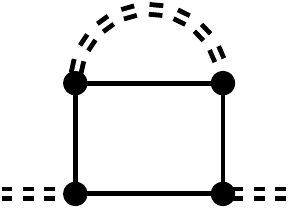}} \;\; \Bigg]_{\Delta (k;\, \phi)\; \approx\; \Delta^{(0)}(k;\, \phi)}
\end{align*}
\caption{Diagrammatic representation of topologies of graphs that are resummed  by means of the EoMs stated in~\eqref{eq:eomsren}. Notice that the very last two-loop self-energy diagram is Two-Particle-Reducible with respect to the fermion lines and needs be consistently considered in this 
semi-perturbative  treatment of fermion quantum effects.\label{fig:EoM}}
\end{figure}

\begin{figure}[t]
\centering
\includegraphics[width=0.95\textwidth]{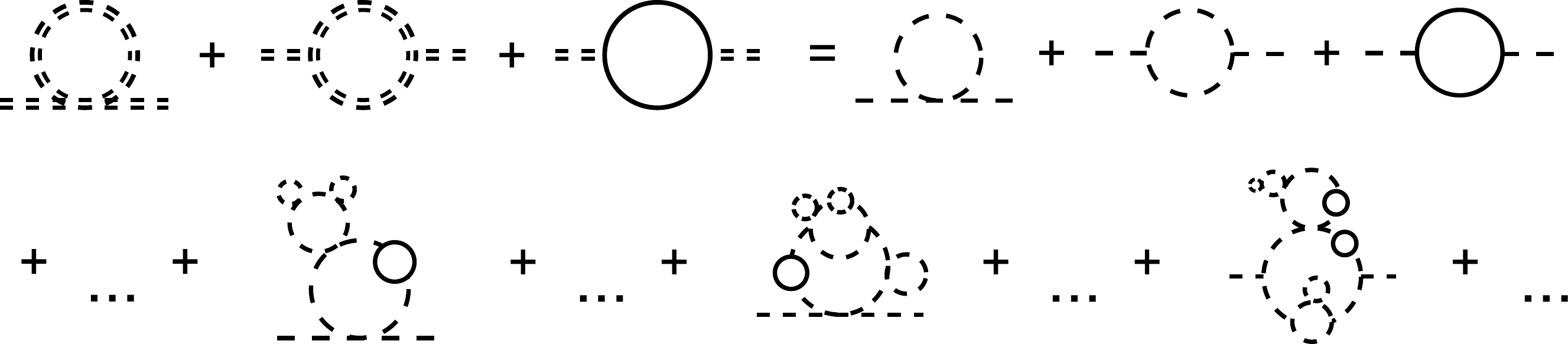}
\caption{Typical topologies of graphs that are implicitly resummed by the one-loop 2PI self-energies in the first line of Figure~\ref{fig:EoM}. Observe that the fermion propagators do not get dressed and are treated perturbatively as in the 1PI formalism.\label{fig:resum_1loop}}
\end{figure}

\begin{figure}[t]
\centering
\includegraphics[width=0.85\textwidth]{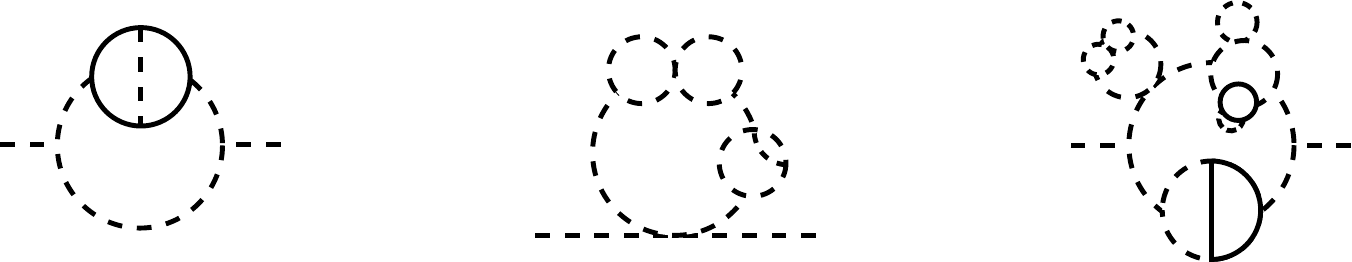}
\caption{Typical set of graphs that are resummed when including the two-loop 2PI self-energies in the second and third line of Figure~\ref{fig:EoM}. Notice that the propagators belonging to two-loop 2PI topologies do not get dressed, in the assumed 
approximation $\Delta(k;\,\phi) \approx \Delta^{(0)}(k,\,\phi)$.\label{fig:resum_2loop}}
\end{figure}

It is now instructive to understand diagrammatically the full set of topologies of graphs that are resummed by means of the EoMs for the dressed  propagators denoted collectively as~$\Delta(k;\,\phi) \equiv \Delta^{H,\,G,\,+}(k;\,\phi)$.  In fact, Figure~\ref{fig:EoM} shows graphically the sort of diagrams that are accounted for by the EoMs stated in~\eqref{eq:eomsren}.  The first three graphs in the first line of Figure~\ref{fig:EoM} represent an infinite set of diagrams of certain topologies, as shown more explicitly in Figure~\ref{fig:resum_1loop}. In order to be able to compare our results with the perturbative two-loop calculation of the SM effective potential, we have also included the two-loop 2PI diagrams in the second and third lines of Figure~\ref{fig:EoM}. In this way, we may also resum diagrams as the ones depicted in Figure~\ref{fig:resum_2loop}.  Since we approximate in the EoMs~\eqref{eq:eomsren} the dressed scalar propagators $\Delta(k;\,\phi)$ appearing in these two-loop self-energies with their respective tree-level forms $\Delta^{(0)} (k;\,\phi) \equiv \Delta^{(0),H,\,G,\,+}(k;\,\phi)$, the corresponding lines in Figure~\ref{fig:resum_2loop} do not get dressed. The same is true for the fermion lines, which represent tree-level propagators as in the 1PI formalism. In summary, we take into account the full contribution of two-loop diagrams and, in addition, a much larger class of diagrams, as compared to the approximate resummation method outlined in Section~\ref{sec:IR_pert}. Unlike in the latter method, the momentum dependence of all the resummed graphs is retained in the SI2PI approach.

At this point, it is important to emphasize that in the SI2PI approach, Goldstone-boson IR divergences are absent by construction. Such divergences could only occur, if two or more Goldstone-boson propagators carry the same momentum, as in the ring diagrams shown in Figure~\ref{fig:IR}. Specifically, the IR divergences originate from a series of Goldstone-boson self-energies occurring in single Goldstone-boson lines. However, such topologies are necessarily \emph{Two-Particle-Reducible} (2PR), and as such,  they do not appear in the diagrammatic series of~$\Gamma[\phi,\Delta]$, which contains only 2PI diagrams with respect to the scalar-field propagators. Hence, the resummation of all Goldstone-boson IR divergences is automatic in the SI2PI formalism and can thus be performed in a systematic manner. 

\vfill\eject

As presented in~\cite{Pilaftsis:2013xna}, the effective potential in the SI2PI formalism may be computed by means of the standard 1PI Ward identity 
\begin{equation}
   \label{eq:WI_phi}
 - \, \frac{d \widetilde
  V_{\rm{eff}}(\phi)}{d \phi} \ \equiv \ \phi \, \Delta^{-1,\,G}(k=0;\phi) \;.
\end{equation}
In fact, the differential equation~\eqref{eq:WI_phi} should be viewed as a fundamental equation whose solution \emph{defines} the SI2PI effective potential, which we denote as $\widetilde V_{\rm{eff}}(\phi)$ so as to dis\-tinguish it from the usual 1PI effective potential~$V_{\rm{eff}}(\phi)$, frequently used in the literature. In the custodial $SU(2)_C$ symmetric limit of the theory, such a definition is unique, because of the equality of the resummed neutral and charged Goldstone-boson propagators, i.e.~$\Delta^{G}(k;\phi) = \Delta^{+}(k;\phi)$. However, as discussed in detail in Section~\ref{sec:ferm}, the inclusion of chiral fermion quantum effects breaks explicitly this custodial symmetry, leading to a potential ambiguity, since one might have used $\Delta^{-1,+}(k=0;\phi)$ on the RHS of~\eqref{eq:WI_phi} to define~$\widetilde{V}_{\rm{eff}}$. Although the finite part of the CT $\delta \lambda^{\rm cb}_1$ is chosen so as to match the two resummed propagators at the minimum of the SI2PI effective potential, i.e.
\begin{equation}
\Delta^{-1,\,G}(k=0;\, v)\ =\ \Delta^{-1,\,+}(k=0;\, v)\ =\ 0\;,
\end{equation}
far away from the minimum, e.g.~for~$\phi \gg v$, the two versions of the so-derived effective potentials may slightly differ from each other, through higher-order effects. The origin of this small difference is due to the scheme  assumed to renormalize the aforementioned spurious custodially violating effects in the EoM of~$\Delta^{+}(k;\phi)$, instead of~$\Delta^G(k;\phi)$. Nevertheless, we find that the numerical impact of these renormalization scheme-dependent effects is negligible for the purposes of this study.

We utilize the computational method developed in~\cite{Pilaftsis:2013xna} to solve numerically 
the EoMs given in~\eqref{eq:eomsren}, for different values of $\phi$,  The numerical solution for~$\Delta^G(k;\phi)$ is then employed to evaluate the SI2PI effective potential, by means of~\eqref{eq:WI_phi}. The boundary condition of the differential equation~\eqref{eq:WI_phi} may be chosen such that $\widetilde{V}_{\rm eff} = 0$ at its minimum $\phi=v$. As a non-trivial check, we have expanded the EoMs through two-loop order and have reproduced numerically the well-known perturbative two-loop results in the literature~\cite{Ford:1992pn}. This cross check reassures the correctness of the renormalization procedure presented in Section~\ref{sec:ferm} and in~\ref{app:ren}, and establishes a high degree of accuracy for our
numerical method.

\begin{figure}[t]
\centering
\includegraphics[width=0.9\textwidth]{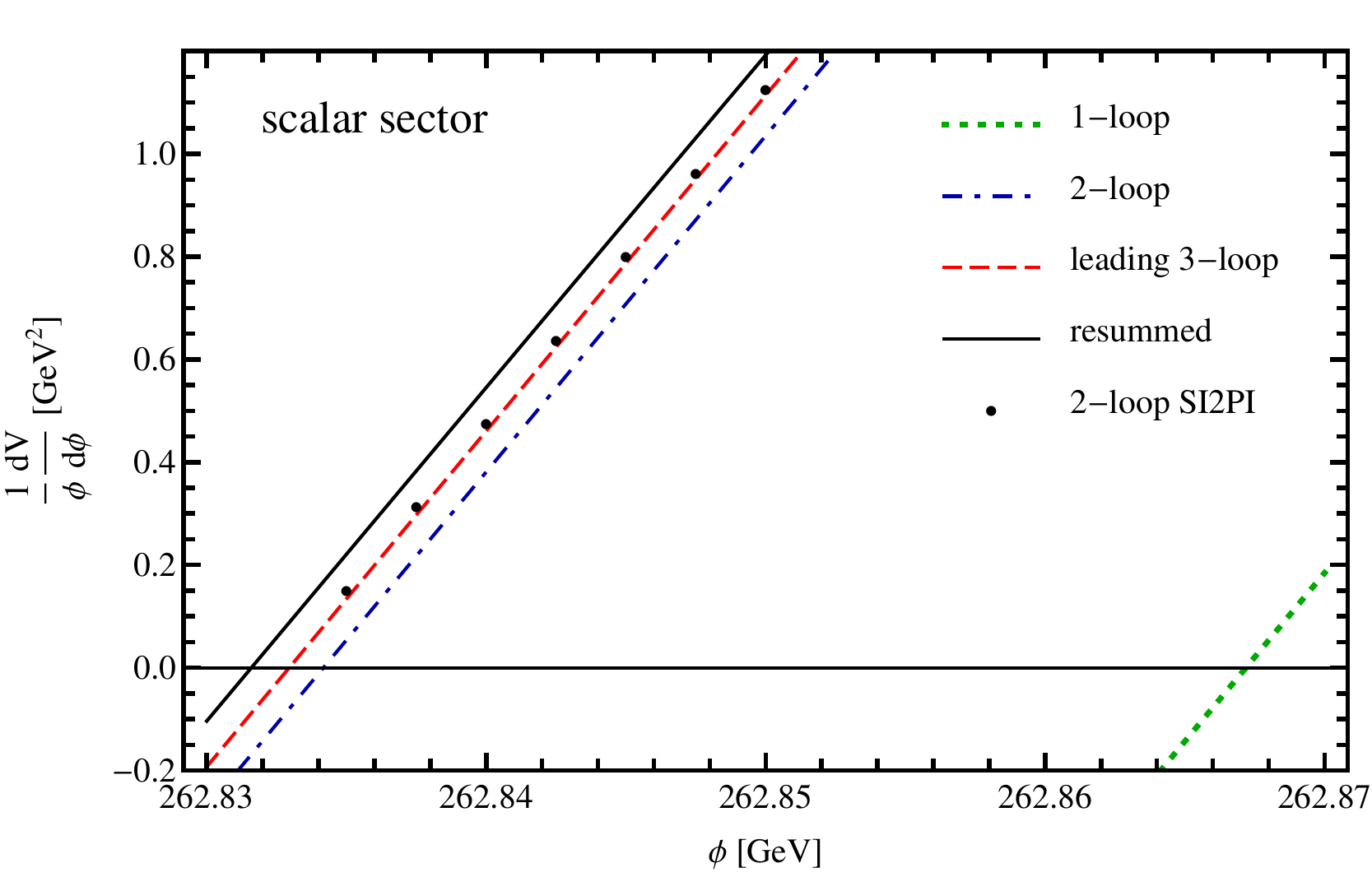}
\caption{Numerical estimates of $(1/\phi)\, dV_{\rm eff}/d\phi$, as a function of~$\phi$ in the vicinity of the dressed minimum $\phi=v$, located at values of $\phi \approx 262.8$~GeV, including {\em only} scalar quantum effects.  Predictions for the perturbative one-loop, two-loop and leading three-loop computations are given by the dotted (green), dash-dotted (blue) and dashed (red) lines, respectively. The solid (black) line is the result of the approximate partial resummation procedure discussed in Section~\ref{sec:IR_pert}. The black dots are the results obtained in the SI2PI 
approach. The same input parameters as in Figure~\ref{fig:IR_pert} are used.\label{fig:SI2PI_scalar}}
\end{figure}
 
Let us first discuss the results obtained by considering only scalar quantum effects on the effective potential in the gaugeless SM. In Figure~\ref{fig:SI2PI_scalar}, we present numerical estimates of the quantity $(1/\phi)\, dV_{\rm eff}/d\phi$, as a function of~$\phi$ in the vicinity of the dressed minimum $\phi=v$, which is located at values of $\phi \approx 262.8$~GeV.  The predictions for the perturbative one-loop, two-loop and leading three-loop computations are given by the dotted (green), dash-dotted (blue) and dashed (red) lines, respectively.  We note that the black dots in Figure~\ref{fig:SI2PI_scalar} represent the numerical solution obtained in the SI2PI approach, which should be contrasted with the solid (black) line displaying the results obtained with the approximate
partial resummation method outlined in Section~\ref{sec:IR_pert}. The difference between the two different approaches is significant, as it is about 75\% of the sum of three- and higher-loop contributions to the effective potential. In this respect, we observe that the leading three-loop result is larger than the naive expectation, which is about $\lambda/16 \pi^2$ times the two-loop one. This fact indicates a breakdown of finite-order perturbation theory, as discussed in Section~\ref{sec:IR}. 
Thus, resumming IR-enhanced contributions in a complete and self-consistent manner, as done in the SI2PI approach, becomes an essential and indispensable task in higher-order precision computations of the SM effective potential.

\begin{figure}[t]
\centering
\includegraphics[width=0.6\textwidth]{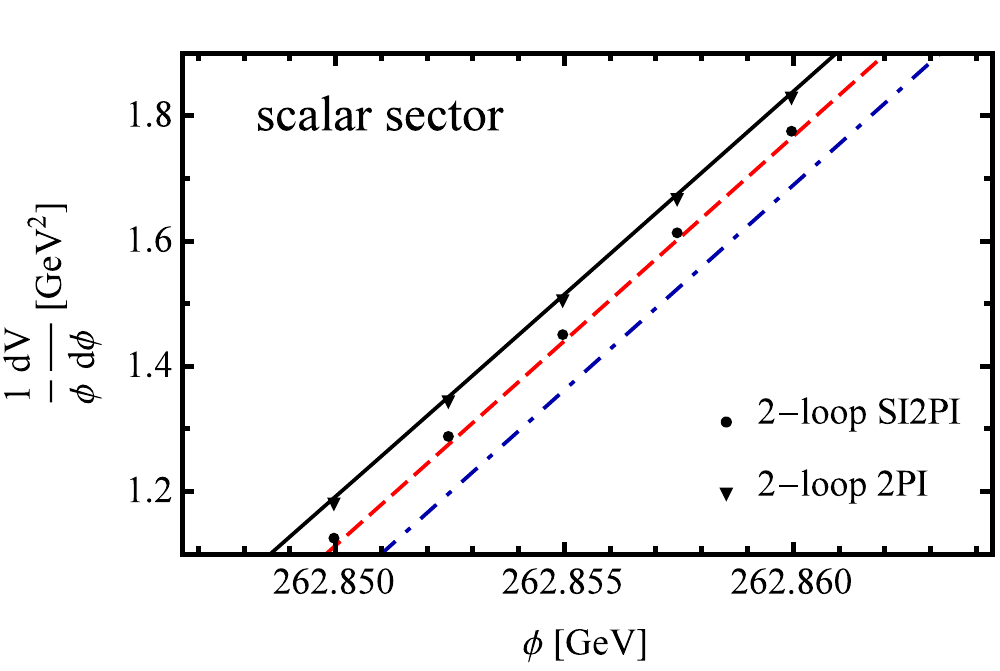}
\caption{Comparison of the results for $(1/\phi)\, dV_{\rm eff}/d\phi$ obtained in the SI2PI (black dots) and 2PI (black triangles) formalisms. The lines and the input parameters are the same as those in Figure~\ref{fig:SI2PI_scalar}.\label{fig:2PIvsSI2PI}}
\end{figure}

It is now interesting to assess the significance of our results obtained in the SI2PI formalism by comparing them with those derived in the conventional 2PI framework.  To~this~end, we truncate the standard 2PI effective action at the two-loop order and compute the derivative of the effective potential $d V_{\rm eff}/d \phi$, in Euclidean space, as follows: 
\begin{align}
   \label{eq:standard2PI}
\frac{d V_{\rm eff}}{d \phi} \ &\equiv \ \frac{1}{\mathrm{V}_4} \, \frac{\delta \Gamma[\phi,\Delta(\phi)]}{\delta \phi} \ = \ \frac{1}{\mathrm{V}_4} \, \frac{\delta \Gamma[\phi,\Delta]}{\delta \phi} \bigg|_{\Delta(\phi)} \notag \\
&= \ \phi \, \big( \,\lambda \phi^2 \,-\, m^2  \:+\: 3
                         \lambda  \, \mathcal{T}_H \:+\: 3 \lambda \,\mathcal{T}_G \,\big) \: 
+ \: T^{(2)}_H \;,
\end{align}
where $\mathrm{V}_4$ is the infinite 4-volume of integration, $\mathcal{T}_{H,G}$ are the renormalized Higgs- and Goldstone-boson integrals [cf.~\eqref{eq:eomsren_scal}], and $T^{(2)}_H$ is the two-loop tadpole contribution reported in \ref{app:int}. Note that the latter has consistently been approximated by its perturbative 1PI form.  In the standard 2PI framework, the propagators are evaluated using the EoMs in~\eqref{eq:eomsren_scal}, but without including the two-loop self-energies $\Pi^{\mathrm{2PI}, (2)}_{H,G}$. Thus, there is a mismatch through the different loop order that the EoMs for the propagators and for the fields have been truncated, which affects the predictions for the effective potential in the standard 2PI formalism.  From Figure~\ref{fig:2PIvsSI2PI}, it is obvious that the discrepancy in the predictions obtained between the two methods is numerically significant. Specifically, the results derived by the standard 2PI formalism are quantitatively close to the ones found in the approximate partial resummation method discussed above. Consequently, this exercise demonstrates that the SI2PI formalism provides a unique and consistent framework for resumming IR-enhanced contributions to the effective potential.  Finally, we note that the solution pertaining to the Goldstone-boson propagator in the standard 2PI formalism becomes tachyonic for $\phi \lesssim 262.85 \, {\rm GeV}$, so that its naive use in the vicinity of the dressed minimum of the effective potential turns out to be problematic. Again, this feature is absent in the SI2PI approach, where the Goldstone-boson propagator becomes tachyonic, by construction, only for field values $\phi$ smaller than the minimum of $\widetilde{V}_{\rm eff}(\phi)$~[cf.~\eqref{eq:WI_phi}].

\begin{figure}[t]
\centering
\includegraphics[width=0.9\textwidth]{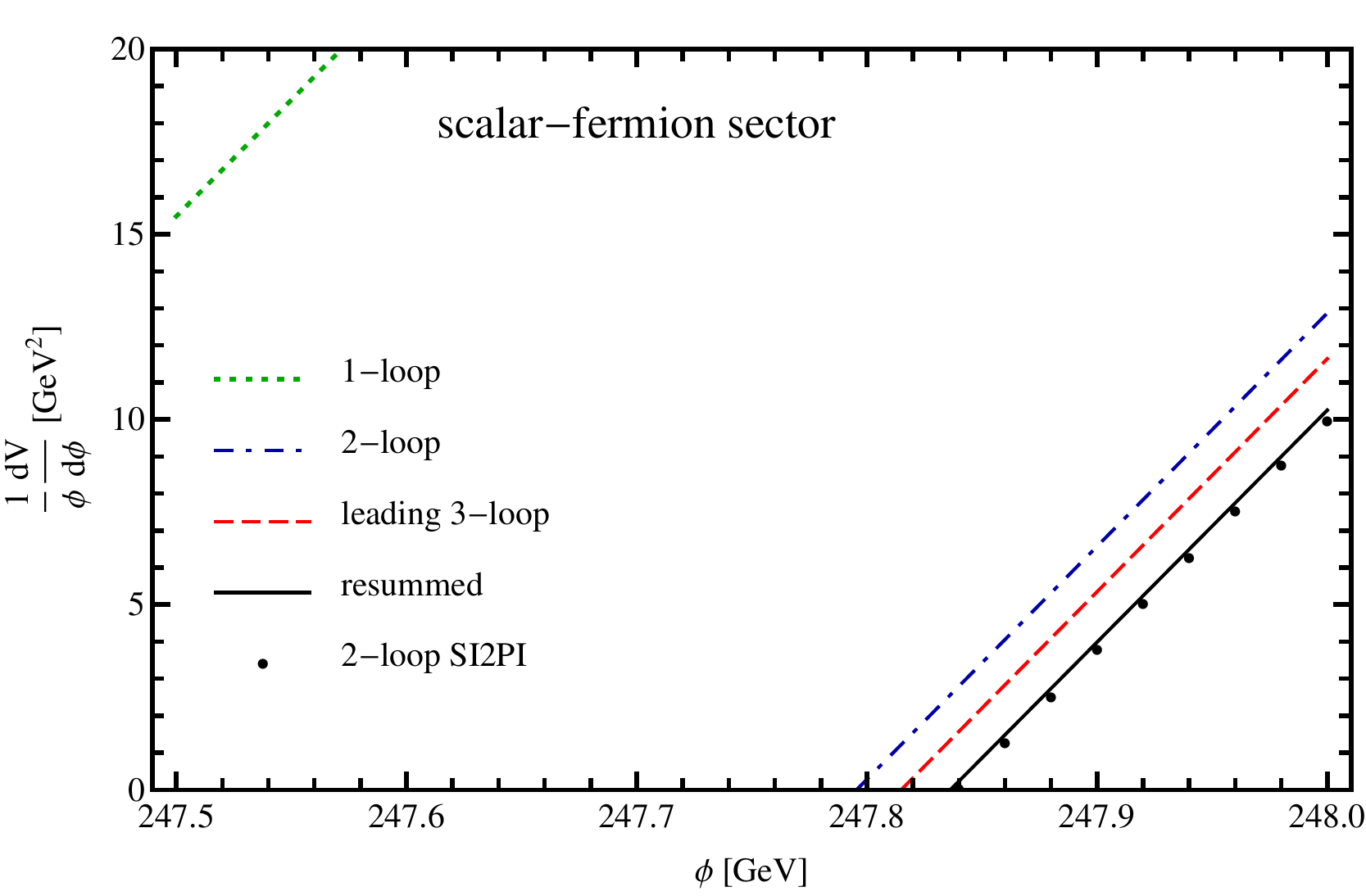}
\caption{The same as in Figure~\ref{fig:SI2PI_scalar}, after including both scalar and chiral fermion quantum effects. The value of the top Yukawa coupling $h_t = 0.93697$ is used in the $\overline{\rm MS}$ scheme. The remaining input parameters are chosen as in Figure~\ref{fig:IR_pert}. \label{fig:SI2PI_scalar_fermion}}
\end{figure}

We now turn our attention to the contribution of mixed scalar-fermion quantum effects to the SM effective potential in the gaugeless limit of the theory. In Figure~\ref{fig:SI2PI_scalar_fermion}, we show numerical estimates for $(1/\phi)\, dV_{\rm eff}/d\phi$, as a function of~$\phi$, by solving the complete EoMs~\eqref{eq:eomsren}, in which the contributions of fermion quantum loops are included.  In this case, we find that the results obtained in the SI2PI approach are in fair agreement with those found in the approximate resummation method of~\cite{Martin:2014bca, Elias-Miro:2014pca}.

\begin{figure}[t]
\centering
\includegraphics[height=5.5cm]{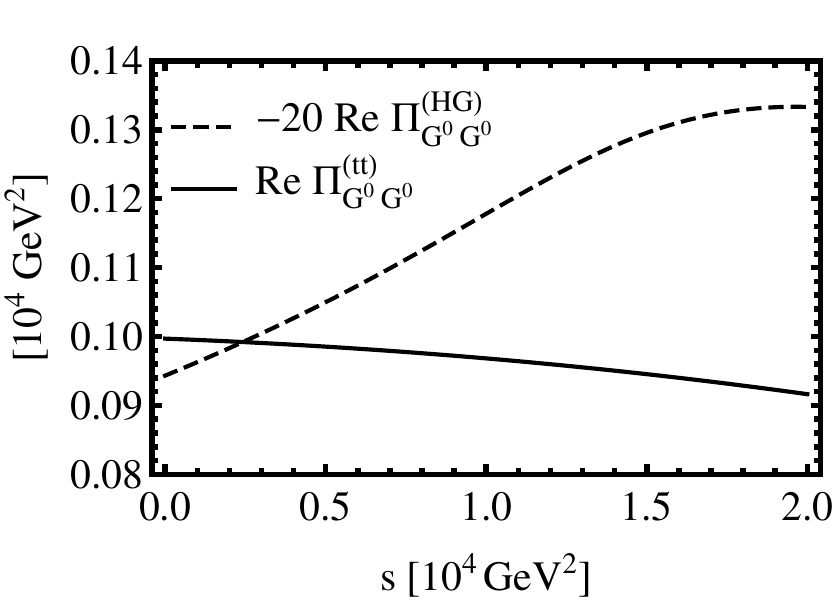}
\caption{Numerical estimates of the real part of 1PI two-loop self-energies of neutral Goldstone bosons involving quantum corrections due to only scalars {\em or} chiral fermions, as functions of the variable $s$. Observe that the momentum dependence of the latter quantum corrections in the IR region is much weaker. \label{fig:self_compar}}
\end{figure}

\begin{figure}[t] \centering \includegraphics[height=5.5cm]{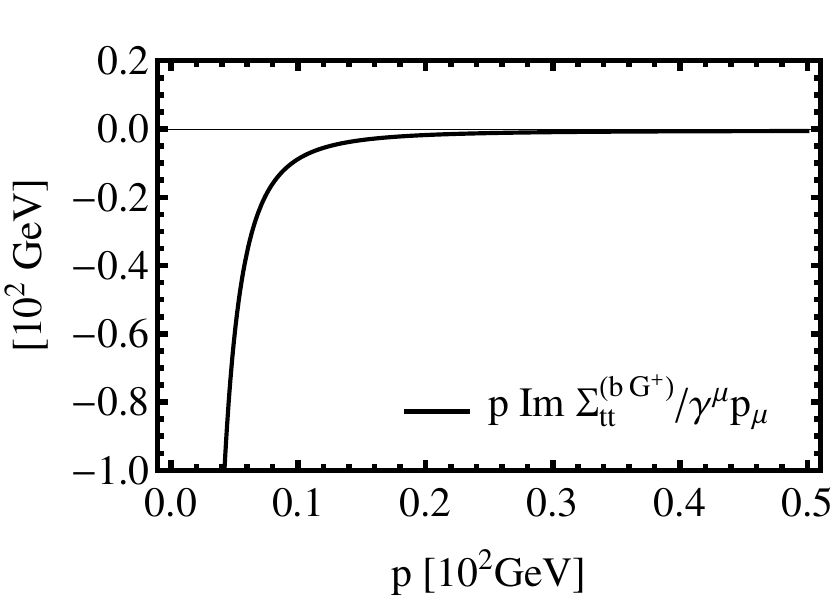}\rule{0.05cm}{0cm} \caption{Numerical values of the expression $p\, {\rm Im}\, \Sigma_{tt}^{(b\,G^+)}(\!\not\! p)/\!\!\not\!p$, which is related to the absorptive part of the 1PI one-loop self-energy of the $t$-quark induced by the on-shell contribution from a $b$-quark and a $G^+$ boson in the loop, as a function of $p\equiv \sqrt{s}$.  This illustrates that the momentum dependence of the dressed top-quark propagator gets significantly modified in the deep IR region, with respect to the tree-level one.\label{fig:top_dressing}}
\end{figure}

This last result is not obvious, but it can be understood in terms of the momentum-dependence of the Goldstone-boson self-energies. As shown in Figure~\ref{fig:self_compar}, fermion quantum effects yield the largest  contribution to the Goldstone-boson self-energies, and unlike scalar quantum effects, they show a weaker momentum dependence, which is almost constant in the~IR.  Since the approximate partial resummation neglects the momentum-dependence of the resummed self-energies, it works significantly better in the mixed scalar-fermion case, as the latter is dominated by the top-quark contribution thanks to the large top Yukawa coupling $h_t$. However, in a full 2PI analysis, resummation effects may potentially alter this conclusion. As shown in Figure~\ref{fig:top_dressing}, multi-particle threshold effects from the dressed top-quark propagator could significantly modify the momentum dependence of the Goldstone-boson propagators in the relevant IR region, and so they can give rise to possible sizeable deviations from the predictions derived with the approximate partial resummation method mentioned above. A detailed discussion of these effects in a full 2PI approach goes beyond the scope of this work and may be given elsewhere.

\vfill\eject

\section{Conclusions}\label{sec:conclusions}

The 2PI effective action constitutes a first-principles systematic approach to consistently resum infinite series of selected sets of diagrams. In this selective resummation approach, one does not run into the risk of over-counting graphs, and so no {\it ad hoc} subtractions are needed to achieve single counting. In its symmetry-improved version considered here, the SI2PI formalism is a rigorous framework to study models with global symmetries. In this paper we have applied this formalism to the SM in the gaugeless limit of the theory, namely to an electroweak model realizing a global $SU(2)_L \times U(1)_Y$ group with vanishing gauge couplings. Specifically, the field content of the gaugeless SM that we have been studying consisted of one Higgs doublet, one left-handed top and bottom quark, and one right-handed top quark.  For the purpose of this work, we treated all quantum effects due to chiral fermions semi-perturbatively. This means that we have not considered bilocal sources and dressed propagators for all chiral quarks, but only for the scalar fields in the Higgs doublet, including the ones for the Goldstone bosons.

In this simplified framework of the SI2PI formalism, we have studied the problem of IR divergences of the SM  effective potential due to massless Goldstone bosons related to the longitudinal polarizations of the $W^\pm$ and $Z$ bosons. To this end, we have taken into account all relevant counterterms related to the renormalization of the 2PI effective action, as well as those associated with the renormalization of spurious custodially breaking effects triggered by the top Yukawa couplings. We have calculated the SI2PI effective potential and have shown that it is IR finite, thereby providing a firm proof of earlier observations that the IR divergences in the 1PI effective potential are an artifact of perturbation theory. This conclusion is not only valid for the ungauged SM under study, but {\em general}. It applies to the full SM, as well as to models of New Physics realizing SSB of extra gauge groups.

The results of our analysis have been compared with those derived from an approximate partial resummation of Goldstone-boson ring diagrams, as done so far in the literature to address the Goldstone-boson IR problem. Moreover, we have compared our results with those that would have been found, if we had calculated the effective potential within the standard 2PI framework.  By considering {\em only} quantum scalar effects of the SM, we have shown that the results obtained in our SI2PI approach differ in a relevant manner with the ones predicted by the aforementioned approximate resummation method {\em and} the standard 2PI approach. Specifically, the difference in the predictions between the SI2PI and the other two methods was found to be numerically significant, i.e.~about~75\% of the sum of three- and higher-loop contributions.

This sizeable difference in the predictions is not generic, but alters considerably,  once the contributions from top- and bottom-quark loops were added. In this case, we have found fairly good agreement with the previously quoted estimates of the approximate method.  The latter is not an obvious result and may be partially attributed to the kinematic behaviour of the Goldstone-boson self-energies in the IR region. Fermion quantum effects give the biggest contribution to the Goldstone-boson self-energies, and unlike scalar quantum effects, they show a weaker momentum dependence, which is almost constant in the~IR.  Nevertheless, we have argued that such a conclusion may be premature and can potentially alter in a full 2PI analysis.  We have demonstrated that if the fermions are treated non-perturbatively within the SI2PI formalism as well, multi-particle threshold effects can modify significantly the momentum dependence of the Goldstone-boson propagators in the relevant IR region, thus leading to possible significant deviations from the predictions found using the approximate partial resummation method mentioned above.  A detailed discussion of the fully non-perturbative inclusion of chiral fermions in the gaugeless SM lies beyond the scope of this work and may be given elsewhere.

We note that the SI2PI approach developed further in this paper can find an immediate application to precision computations of the effective potential in supersymmetric extensions of the SM.  In particular, in the so-called Minimal Supersymmetric Standard Model~(MSSM), scalar top quarks provide the dominant source of radiative corrections~\cite{ERZ,HH,OYY}, in which case the aforementioned approximate partial resummation method is bound to be inadequate to deal with the higher-order precision required in the computation.  Instead, in view of the present study, the SI2PI formalism proves itself to be a rigorous and systematic approach that allows one to accurately address the Goldstone-boson IR problem in the MSSM effective potential. Finally, the SI2PI formalism may be used to study, from first principles and with higher precision, IR-sensitive renormalon effects in Quantum Chromo\-dynamics~\cite{Beneke:1998ui,Dasgupta:1996hh}. It~would be interesting to report progress on the above issues in the near future.

\section*{Acknowledgements}
\vspace{-3mm}
\noindent
The work of A.P. is supported by the Lancaster-Manchester-Sheffield Consortium for Fundamental Physics under STFC grant ST/L000520/1. The work of D.T. is funded by the Belgian Federal Science Policy through the Interuniversity Attraction Pole P7/37.

\newpage

\appendix

\makeatletter
\gdef\thesection{\appendixname~\@Alph\c@section}%
\makeatother

\section{Loop integrals}\label{app:int}

In this appendix, we present analytical formulae for the renormalized two-loop self-energies $\Pi^{\mathrm{2PI}, (2)}_a$ and $\Sigma^{\mathrm{2PI}, (2)}_a$ appearing in the EoMs ~\eqref{eq:eomsren_scal} and~\eqref{eq:eomsren}. These are calculated in perturbation theory by standard techniques~\cite{Martin:2003it, Martin:2003qz}. As discussed in Sections~\ref{sec:scalar} and~\ref{sec:ferm}, we approximate them by their zero-momentum value. This simplification introduces an  error that is negligible for the purposes of this work. We adopt the compact notation of~\cite{Martin:2003it}, and introduce  $\overline{\ln} x \equiv \ln(x/\mu^2)$ and $s=-k^2$, where the momentum $k$ is given in the Wick-rotated Euclidean space. We may now define the one-loop functions
\begin{subequations}
\begin{align}
A(x) \ &\equiv \ x \, (\overline{\ln} x - 1) \;,\\
B(k;x,y) \ &\equiv \ - \int_0^1 \! dt \, \overline{\ln}[t x + (1-t) y - t(1-t)s] \;.
\end{align}
\end{subequations}
For a concise presentation of our analytic results, we may also need to define the two-loop functions $I(x,y,z)$, $V(x,y,z,w)$ and $U(x,y,z,w)$, evaluated at zero momentum. The analytical formulae for these loop functions, together with the explicit expression of $B(k;x,y)$, may be found in~\cite{Martin:2003qz}.

For the 2PI self-energies involving only scalar loops, the relevant topologies are the ones depicted in the second line of Figure~\ref{fig:EoM}. Introducing the tree-level background masses squared $h= 3 \lambda \phi^2 - m^2$ and $g = \lambda \phi^2 - m^2$, we find~\cite{Pilaftsis:2015cka}
\begin{subequations}
\begin{align}
(16 \pi^2)^2\, \Pi^{\mathrm{2PI}, (2)}_{H}(\phi) \ &= \ 54 \lambda^3 \phi^2 \, \overline{\ln}^2 h \: + \: 36 \lambda^3 \phi^2 \, \overline{\ln}\, h \, \overline{\ln}\, g \:+\: 30 \lambda^3 \phi^2 \, \overline{\ln}^2 g \notag \\
&- \ 6 \lambda^2 \, I(h,h,h) \:- \: 6 \lambda^2 \, I(h,g,g) \notag\\
&- \ 216 \lambda^3 \phi^2 \, I(h',h,h) \:-\: 72 \lambda^3 \phi^2 \, I(h',g,g) \: - \: 24 \lambda^3 \phi^2 \, I(g',g,h) \notag\\
&- \ 648 \lambda^4 \phi^4 \, I(h',h',h) \:-\: 144 \lambda^4 \phi^4 \, I(h',g',g) \: - \: 24 \lambda^4 \phi^4 \, I(g',g',h) \;, \displaybreak[0]\\
(16 \pi^2)^2\, \Pi^{\mathrm{2PI}, (2)}_{G}(\phi) \ &= \ 8 \lambda^3 \phi^2 \, B(g,h)^2  \: - \: 24 \lambda^2 I(h,h,h) \:+\: 22 \lambda^2 I(g,h,h) \notag \\
& -\: 16 \lambda^2 I(g,g,h) \:+\: 6 \lambda^2 I(g,g,g) \;,
\end{align}
\end{subequations}
where $B(x,y)$ (without explicitly displaying the momentum argument) is understood to be evaluated at zero momentum and a primed argument denotes a derivative of the loop function with respect to that  argument, e.g. 
\begin{equation}
I(h',h,h) \ \equiv \ \frac{d I(x,h,h)}{d x} \, \bigg|_{x=h} \;.
\end{equation}

\begin{figure}
\centering
$\parbox{3.4cm}{\includegraphics[width=3.4cm]{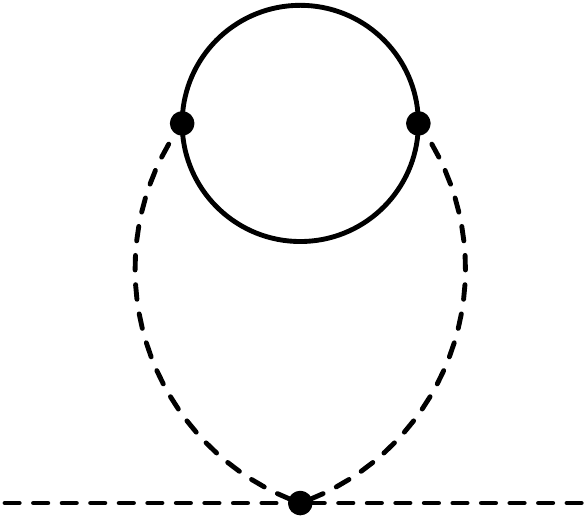}} \hspace{2cm} \parbox{3.4cm}{\includegraphics[width=3.4cm]{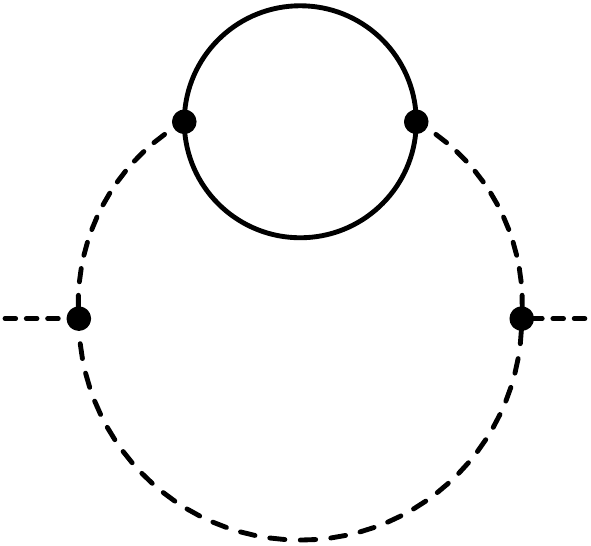}}$
\caption{Perturbative 2PR topologies, with respect to cuts of scalar lines, that contribute to $\Sigma^{\mathrm{2PR}, (2)}_a$.\label{fig:2PR}}
\end{figure}

For the self-energies that include fermion loops, the relevant topologies are the ones shown in the third line of Figure~\ref{fig:EoM}. It proves convenient to calculate first at zero momentum the two-loop self-energies $\Sigma^{\mathrm{2PR}, (2)}_a$ (with $a = H,\,G,\,+$), which are 2PR with respect to cuts of scalar lines, as can be seen from  Figure~\ref{fig:2PR}. Then, we may obtain the 2PI self-energies through the relations
\begin{subequations}\label{eq:relat}
\begin{align}
\frac{1}{\phi} \, \frac{d V^{\rm Yuk,(2)}_{\rm eff}}{d \phi} \ &= \ \Sigma^{\mathrm{2PI}, (2)}_G \; + \; \Sigma^{\mathrm{2PR}, (2)}_G \ = \ \Sigma^{\mathrm{2PI}, (2)}_+ \; + \; \Sigma^{\mathrm{2PR}, (2)}_+ \;,\\
\frac{d^2 V^{\rm Yuk,(2)}_{\rm eff}}{d \phi^2} \ &= \ \Sigma^{\mathrm{2PI}, (2)}_H \; + \; \Sigma^{\mathrm{2PR}, (2)}_H \;.
\end{align}
\end{subequations}
The first relation stems from the standard WI of the 1PI effective action. The well-known result for the perturbative two-loop contribution to the effective potential $V^{\rm Yuk,(2)}_{\rm eff}(\phi)$ thanks to top-quark Yukawa interactions is given by
\begin{align}\label{eq:Yuk_V}
V^{\rm Yuk,(2)}_{\rm eff}(\phi) \ &= \ \frac{3 \, h_t^2}{2\,(16 \pi^2)^2} \, \Big[ 2 A(t)^2 \, - \, 4 A(t) A(g) \, - \, 2 A(t) A(h) \, + \, (4 t - h) I(t,t,h) \notag\\
& +\ 2 (t - g) I(t,g,0) \, - \, g \, I(t,t,g) \Big] \;.
\end{align}
The 2PR self-energies $\Sigma^{\mathrm{2PR}, (2)}_a$ are calculated to be
\begin{subequations}
\begin{align}
(16 \pi^2)^2\, \Sigma^{\mathrm{2PR}, (2)}_{H}(\phi) \ = \ &- \, 9 \lambda h_t^2 \,\Big[(h - 4 t) I(h',t,t) \, + \, I(h,t,t) \, + \, 2 \, A(t) \, \overline{\ln} \,h \Big]\notag \\
& - \, 3 \lambda h_t^2 \,\Big[g \, I(g',t,t) \, + \, I(g,t,t) \, + \, 2 \, A(t) \, \overline{\ln} \,g \Big] \notag \\
&- \, 6 \lambda h_t^2 \,\Big[(g- t) \, I(g',t,0) \, + \, I(g,t,0) \, + \,  A(t) \, \overline{\ln} \,g \Big] \notag \\
&+ \, 108 \lambda^2 h_t^2 \phi^2 \, \Big[ (4 t - h)  V(h,h,t,t) \, + \, U(h,h,t,t) \, +\, 2 A(t) B(h',h) \Big] \notag\\
&+ \, 12 \lambda^2 h_t^2 \phi^2 \, \Big[ - \,g \,  V(g,g,t,t) \, + \, U(g,g,t,t) \, +\, 2 A(t) B(g',g) \Big] \notag\\
&+ \, 24 \lambda^2 h_t^2 \phi^2 \, \Big[ (t-g) V(g,g,t,0) \, + \, U(g,g,t,0) \, +\, A(t) B(g',g) \Big] \;,\displaybreak[0]\\
(16 \pi^2)^2\, \Sigma^{\mathrm{2PR}, (2)}_{G}(\phi) \ = \ &- \, 3 \lambda h_t^2 \,\Big[(h - 4 t) I(h',t,t) \, + \, I(h,t,t) \, + \, 2 \, A(t) \, \overline{\ln} \,h \Big]\notag \\
& - \, 9 \lambda h_t^2 \,\Big[g \, I(g',t,t) \, + \, I(g,t,t) \, + \, 2 \, A(t) \, \overline{\ln} \,g \Big] \notag \\
&- \, 6 \lambda h_t^2 \,\Big[(g- t) \, I(g',t,0) \, + \, I(g,t,0) \, + \,  A(t) \, \overline{\ln} \,g \Big] \notag \\
&+ \, 12 \lambda^2 h_t^2 \phi^2 \, \Big[ (4 t - h)  V(g,h,t,t) \, + \, U(g,h,t,t) \, +\, 2 A(t) B(h',g) \Big] \notag\\
&+ \, 12 \lambda^2 h_t^2 \phi^2 \, \Big[ - \,g \,  V(h,g,t,t) \, + \, U(h,g,t,t) \, +\, 2 A(t) B(g',h) \Big] \;,\displaybreak[0]\\
(16 \pi^2)^2\, \Sigma^{\mathrm{2PR}, (2)}_{+}(\phi) \ = \ &- \, 3 \lambda h_t^2 \,\Big[(h - 4 t) I(h',t,t) \, + \, I(h,t,t) \, + \, 2 \, A(t) \, \overline{\ln} \,h \Big]\notag \\
& - \,3 \lambda h_t^2 \,\Big[g \, I(g',t,t) \, + \, I(g,t,t) \, + \, 2 \, A(t) \, \overline{\ln} \,g \Big] \notag \\
&- \, 12 \lambda h_t^2 \,\Big[(g- t) \, I(g',t,0) \, + \, I(g,t,0) \, + \,  A(t) \, \overline{\ln} \,g \Big] \notag \\
&+ \, 12 \lambda^2 h_t^2 \phi^2 \, \Big[ (4 t - h)  V(g,h,t,t) \, + \, U(g,h,t,t) \, +\, 2 A(t) B(h',g) \Big] \notag\\
&+ \, 12 \lambda^2 h_t^2 \phi^2 \, \Big[ (t - g )  V(h,g,t,0) \, + \, U(h,g,t,0) \, +\, A(t) B(g',h) \Big] \;.
\end{align}
\end{subequations}
As discussed above, the 2PI two-loop self-energies $\Sigma^{\mathrm{2PI}, (2)}_a$ (with $a = H,\,G,\,+$) are obtained by combining these expressions with~\eqref{eq:relat} and~\eqref{eq:Yuk_V}.

Finally, we give the analytical expression for the two-loop tadpole contribution~$T^{(2)}_{H}(\phi)$
to the Higgs field, 
\begin{equation}
(16 \pi^2)^2\, T^{(2)}_{H}(\phi) \ = \ - \ 6 \lambda^2 \, I(h,h,h) \:- \: 6 \lambda^2 \, I(h,g,g)\;.
\end{equation}
This last expression will be needed for the evaluation in~\eqref{eq:standard2PI} of the derivative of the effective potential~$dV_{\rm eff}/d\phi$ in the standard 2PI formalim.

\section{Renormalization of Mixed Scalar-Fermion Quantum Loops}\label{app:ren}

In this appendix, we complete the discussion of Section~\ref{sec:ferm} on the renormalization of the EoMs  stated in~\eqref{eq:eomsbare}, in which both scalar and chiral fermion quantum effects are considered.

We start by renormalizing the tadpole integral $\bm{\mathcal{T}}_{\!\!a}$, along the lines of what was done in Section~\ref{sec:ferm} for the sunset diagrams. To this end, we decompose the scalar dressed propagators as
\begin{equation}
\Delta^{-1,a}(k) \ \equiv \ k^2 \; + \; M_a^2 \;+\; \Pi^{\rm scalar}_a(k) \;+\; \Sigma_a(k) \;,
\end{equation}
where the mass parameter $M^2_a$ contains all momentum-independent terms and $\Pi^{\rm scalar}_a(k)$ stands for  the renormalized scalar sunset diagrams, with $a = H,\,G,\,+$. In terms of the auxiliar propagator~$\widetilde{\Delta}_0(k)$ introduced in~\eqref{eq:aux}, we may perform the asymptotic expansion
\begin{equation}
\Delta^a(k) \ = \ \widetilde{\Delta}_0(k) \: - \: 
\widetilde{\Delta}_0(k) \left(M^2_a-\mu^2 + \Pi^{\rm scalar}_a(k) + \Sigma_a(k) - \widetilde{\Sigma}(k) \right) \widetilde{\Delta}_0(k) \: + \: {\cal O}\left(\widetilde{\Delta}_0^3(k)\right)\;.
\end{equation}
In order to extract the UV divergences of  $\bm{\mathcal{T}}_{\!\!a}$ in terms of the auxiliary propagator $\widetilde{\Delta}_0(k)$, we introduce the integrals
\begin{subequations}
\begin{align}
\widetilde{\bm{\mathcal{T}}}_{\!\!0} \ &\equiv \ \overline{\mu}^{2 \epsilon} \int_k \;\widetilde{\Delta}_0(k)\;,\\
\widetilde{\bm{\mathcal{J}}}_{\!\!0} \ &\equiv \ \overline{\mu}^{2 \epsilon} \int_k \;\widetilde{\Delta}_0(k)^2 \, \frac{B(k;\mu^2,\mu^2)}{16 \pi^2}\;.
\end{align}
\end{subequations}
Hence, we have
\begin{equation}
  \label{eq:calTa}
\bm{\mathcal{T}}_{\!\!a} \ = \ \widetilde{\bm{\mathcal{T}}}_{\!\!0}  \;-\;  (M^2_a-\mu^2) \, \widetilde{\bm{\mathcal{I}}}_{0} \;-\; \int_k \Pi^{\rm scalar}_a(k)\,\widetilde{\Delta}_0^2(k) \; - \; \int_k \left(\Sigma_a(k) - \widetilde{\Sigma}(k)\right) \, \widetilde{\Delta}_0^2(k)  \; + \; (\text{finite})\;.
\end{equation}
Like in the pure scalar case, we may replace in the third term on the RHS of~\eqref{eq:calTa} the dressed pro\-pagators contained in $\Pi^{\rm scalar}_a(k)$ with the auxiliary {\em tree-level} propagator $\Delta_0(k)$ of~\eqref{eq:D0}, since the difference so introduced is UV finite (see~\cite{Pilaftsis:2013xna}). In order to match to the perturbative $\overline{\rm MS}$ scheme at two-loop order, we select the CT parts as described in Section~\ref{sec:ferm}. Thus, the $\overline{\rm MS}$-renormalized tadpole integral $\mathcal{T}_a$ 
may be calculated as follows:
\begin{equation}\label{eq:tad_exp}
\bm{\mathcal{T}}_{\!\!a} \ = \ \widetilde{\bm{\mathcal{T}}}_{\!\!\rm CT}  \;-\;  (M^2_a-\mu^2) \, \widetilde{\bm{\mathcal{I}}}_{\rm CT} \;+\; \nu_a \lambda^2 \phi^2 \widetilde{\bm{\mathcal{J}}}_{\! \rm CT} \; - \; \int_k \left(\Sigma_a(k) - \widetilde{\Sigma}(k)\right) \, \widetilde{\Delta}_0^2(k) \, \Big|_{\rm CT} \; + \; \mathcal{T}_a\;,
\end{equation}
where $\nu_H = 24$ and $\nu_G=\nu_+=4$. The UV-infinite CT parts are obtained as in Section~\ref{sec:ferm}, i.e.
\begin{subequations}
   \label{eq:TJCTs}
\begin{align}
\widetilde{\bm{\mathcal{T}}}_{\!\!\rm CT} \ &\equiv \ \widetilde{\bm{\mathcal{T}}}_{\!\!0} \; - \; \widetilde{{\mathcal{T}}}_0\big|^{(2)}_{\rm fin} \;,\\
\widetilde{\bm{\mathcal{J}}}_{\!\rm CT} \ &\equiv \ \widetilde{\bm{\mathcal{J}}}_{\!0} \; - \; \widetilde{{\mathcal{J}}}_0\big|^{(2)}_{\rm fin} \;.
\end{align}
\end{subequations}
In order to write down the fourth term on the RHS of~\eqref{eq:tad_exp} more explicitly, 
we~make use of the asymptotic expansion
\begin{equation}
  \label{eq:asymp_int_tad}
  \Sigma_a(k) - \widetilde{\Sigma}(k) \ = \ - \, \frac{3 \, h_t^2 }{16 \pi^2} \,\Big[ \xi_a t - 4 \mu^2 - (\chi_a t - 2 \mu^2) B(k;\mu^2,\mu^2)\Big] \; + \; O\bigg(\frac{1}{s}\bigg) \; + \; O(\epsilon) \;.\\
\end{equation}
Here, the values of the coefficients are: $\xi_H = \xi_G = 4$, $\xi_+ = 2$ and $\chi_H = 6$, $\chi_G=\chi_+=2$. By virtue of the asymptotic expansion~\eqref{eq:asymp_int_tad}, one is now in a position to isolate the UV divergences from the fourth term on the RHS of~\eqref{eq:tad_exp} as follows:
\begin{equation}
    \label{eq:exp_int_tad}
\int_k \Big(\Sigma_a(k) - \widetilde{\Sigma}(k)\Big) \, \widetilde{\Delta}_0^2(k) \,\Big|_{\rm CT} \ \supset \ - \, \frac{3 \, h_t^2 }{16 \pi^2} \,( \xi_a t - 4 \mu^2) \, \ICT \; + \; 3 h_t^2 (\chi_a t - 2 \mu^2) \, \JCT \;. \\
\end{equation}
There is an additional finite contribution to~\eqref{eq:exp_int_tad} coming from the $O(\epsilon)$ term in~\eqref{eq:asymp_int_tad}. We find convenient to obtain this term by matching directly the LHS and the RHS of~\eqref{eq:tad_exp}, expanded at two-loop order, in the $\overline{\rm MS}$ scheme. This correction term is found to be
\begin{equation}
  - \frac{3 \, h_t^2 }{(16 \pi^2)^2} \, (\xi_a t - 4 \mu^2) \, \frac{1}{2} \ .
\end{equation}
In this way, we finally obtain the $\overline{\rm MS}$-renormalized tadpole integral
\begin{align}\label{eq:tad_ren}
  \mathcal{T}_a \ &= \ \bm{\mathcal{T}}_{\!\!a}  \; - \: \widetilde{\bm{\mathcal{T}}}_{\!\!\rm CT}  \;+\;  \left[M^2_a-\mu^2 -  \frac{3 \, h_t^2 }{16 \pi^2} \, (\xi_a t - 4 \mu^2)\right] \, \widetilde{\bm{\mathcal{I}}}_{\rm CT}\notag\\
                  &+\; \Big[-\nu_a \lambda^2 \phi^2 \;+\; 3 h^2_t (\chi_a t - 2 \mu^2)\Big] \widetilde{\bm{\mathcal{J}}}_{\! \rm CT} \: + \: \frac{3 \, h_t^2 }{(16 \pi^2)^2} \, (\xi_a t - 4 \mu^2) \, \frac{1}{2}\;.
\end{align}
The finite parts of the integrals involving the auxiliary propagators are needed to evaluate~\eqref{eq:tad_ren}, by means of~\eqref{eq:ICT} and~\eqref{eq:TJCTs}. As we did for the loop integrals in~\ref{app:int}, these finite pieces can be calculated in analogous manner. More explicitly, we find
\begin{subequations}
  \begin{align}
    \widetilde{{\mathcal{T}}}_0\big|^{(2)}_{\rm fin} \ &= \Bigg[  \parbox{1.3cm}{\quad\includegraphics[height=1.7cm]{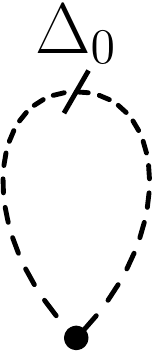}} \;\;+\;\; \parbox{2.2cm}{\quad\includegraphics[height=1.7cm]{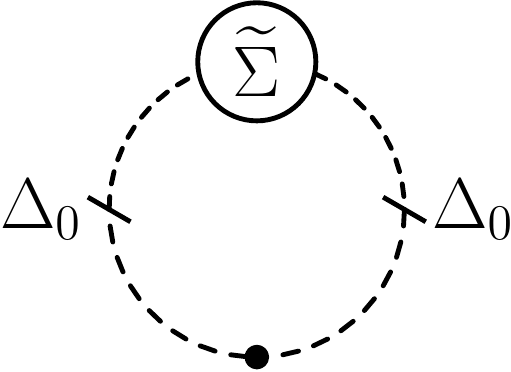}} \quad \Bigg]_{\rm fin} = \ - \, \frac{\mu^2}{16 \pi^2} \left(1 - \frac{3 \, h_t^2}{16 \pi^2} \,\eta_3\right) \;,\\[6pt]
    \widetilde{{\mathcal{I}}}_0\big|^{(2)}_{\rm fin} \ &= \Bigg[  \;\; 2 \;\; \parbox{2cm}{\includegraphics[width=2.1cm]{D0D0SigmatD0}} \quad \Bigg]_{\rm fin} = \ - \, 2\, \frac{3 \, h_t^2}{(16 \pi^2)^2} \,  \eta_2 \;, \label{eq:fin_int_b}\\
    \widetilde{{\mathcal{J}}}_0\big|^{(2)}_{\rm fin} \ &= \Bigg[ \;\;\; \parbox{2.2cm}{\includegraphics[height=2.55cm]{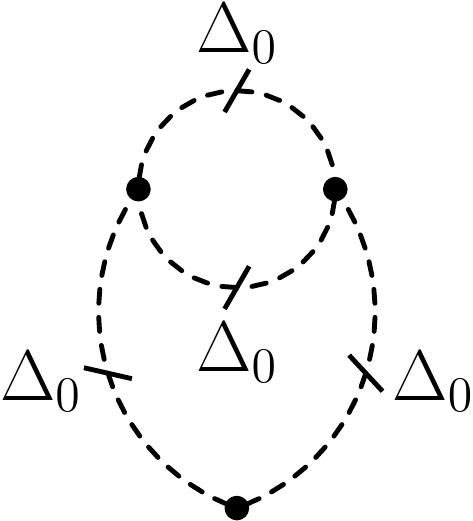}} \;\; - \;\; \parbox{2.4cm}{\includegraphics[height=1.7cm]{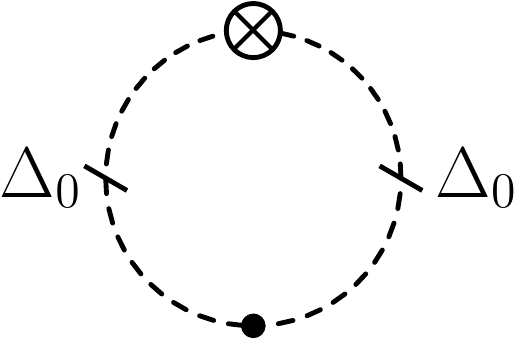}} \quad \Bigg]_{\rm fin}= \ - \, \frac{1}{(16 \pi^2)^2} \, \frac{1}{2} \, \eta_1 \;,
  \end{align}
\end{subequations}
where
\begin{subequations}
  \begin{align}
    \eta_2 \ &= \ 1 \; - \; \frac{2 \, i}{3 \sqrt{3}} \bigg( \mathrm{Li}_2\frac{1 - i \sqrt{3}}{2} - \frac{\pi^2}{36}\bigg) \ \simeq \ 0.60935 \;,\\
    \eta_1 \ &= \ 6\,\eta_2 - 5 \ \simeq \ -1.34391 \;,\\ 
    \eta_3 \ &= \ 12\,\eta_2 - 4 \ \simeq \ 3.31219 \;.
  \end{align}
\end{subequations}

Employing~\eqref{eq:sunset_part} and~\eqref{eq:tad_ren} in the EoMs given in~\eqref{eq:eomsbare}, we can separate the UV-divergent parts from the finite renormalized remainders. The latter give the renormalized EoMs listed in~\eqref{eq:eomsren}. Requiring that  in addition to the wavefunction renormalization, the UV-divergent terms proportional to $\phi^2$, $\mathcal{T}_a$ and the remaining overall divergences do individually vanish, we arrive at the following 6 independent constraining equations:
\begin{subequations}
\begin{align}
0 \ &= \ - 12 \lambda^2 \ICT \;+\; \delta \lambda_2^A (1 - 6 \lambda  \,\ICT) \;+\; \delta \lambda_2^B (2 - 6 \lambda \,\ICT) \;,\displaybreak[0]\\[6pt]
0 \ &= \ - 8 \lambda^2 \ICT \;+\; \delta \lambda_2^A (1 - 6 \lambda  \,\ICT) \;+\; \delta \lambda_2^B (- 2 \lambda  \,\ICT) \;,\displaybreak[0]\\[6pt]
0 \ &= \ \frac{9 \, h_t^4}{16 \pi^2}\frac{1}{\epsilon} \; + \;  \left(\frac{30 \lambda h_t^4}{16 \pi^2} - 36 \lambda^2 \right) \ICT \;+\; (84 \lambda^3 - 36 \lambda h_t^4) \,\JCT \; + \; \delta \lambda_1^A \;+\; \delta \lambda_2^B \notag\\
& + \ \delta \lambda_2^A \left[ \left(\frac{18 h^4_t}{16 \pi^2}  - 6 \lambda \right) \ICT \, + \, (36 \lambda^2 - 18 h_t^4) \, \JCT \right] \notag\\
& + \ \delta \lambda_2^B \left[ \left(\frac{12 h^4_t}{16 \pi^2}  - 6 \lambda \right) \ICT \, + \, (48 \lambda^2 - 18 h_t^4) \, \JCT \right]
\;,\label{eq:div_c}\displaybreak[0]\\
0 \ &= \ \frac{3 \, h_t^4}{16 \pi^2}\frac{1}{\epsilon} \; + \;  \left(\frac{30 \lambda h_t^4}{16 \pi^2} - 12 \lambda^2 \right) \ICT \;+\; (44 \lambda^3 - 24 \lambda h_t^4) \,\JCT \; + \; \delta \lambda_1^A  \notag\\
& + \ \delta \lambda_2^A \left[ \left(\frac{18 h^4_t}{16 \pi^2}  - 6 \lambda \right) \ICT \, + \, (36 \lambda^2 - 18 h_t^4) \, \JCT \right] \notag\\
& + \ \delta \lambda_2^B \left[ \left(\frac{12 h^4_t}{16 \pi^2}  - 2 \lambda \right) \ICT \, + \, (8 \lambda^2 - 6 h_t^4) \, \JCT \right]
\;,\label{eq:div_d}\displaybreak[0]\\
0 \ &= \ \delta \lambda_1^{\rm cb} \; - \; (\lambda + \delta \lambda_2^B) \frac{6 h_t^4}{16 \pi^2} \, \ICT\;,\label{eq:div_e}\displaybreak[0]\\
0 \ &= \ - \delta m_1^2 \;+\; 6 \lambda \TCT \;+\; 6 \lambda \left(m^2 + \mu^2 - \frac{12 h_t^2 \mu^2}{16 \pi^2}\right) \ICT \; +\; 36 \lambda h_t^2 \mu^2 \JCT \notag\\
& + \ \left(2 \delta \lambda_2^A + \delta \lambda_2^B\right) \left[2 \TCT \,+\, 2 \left(m^2 + \mu^2 - \frac{12 h_t^2 \mu^2}{16 \pi^2}\right) \ICT \,+\, 12 h_t^2 \mu^2 \JCT\right]\label{eq:div_f} \;.
\end{align}
\end{subequations}
The first two constraining relations come from the cancellation of the subdivergences proportional to $\mathcal{T}_a$ in the three EoMs given in~\eqref{eq:eomsren}. Equations~\eqref{eq:div_c} and~\eqref{eq:div_d} are obtained from the ones proportional to $\phi^2$ in the EoMs for $\Delta^H(k)$ and $\Delta^G(k)$, respectively, whilst~\eqref{eq:div_e} is the analogous constraining condition for $\Delta^+(k)$, after imposing~\eqref{eq:div_d}. Finally,~\eqref{eq:div_f} is obtained by cancelling the remaining overall divergence in the three EoMs stated in~\eqref{eq:eomsren}. In this way, the following analytic expressions for the CTs are obtained:
\begin{subequations}\label{eq:CTs}
\begin{align}
\delta Z_1 \ &= \ -\,\frac{3 \, h_t^2}{16 \pi^2}\, \frac{1}{\epsilon} \;,\displaybreak[0]\\
\delta m_1^2 \ &= \ \frac{6 \lambda}{1 - 6 \lambda \ICT} \, \bigg[\Big( \TCT + (m^2 + \mu^2)\,\ICT\Big) \; -\; 3 h^2_t \mu^2 \bigg( \frac{4}{16 \pi^2} \ICT - 2 \JCT\bigg) \bigg] \;,\displaybreak[0]\\
\delta \lambda_1^A \ &= \ - \, \frac{3\,h_t^4}{16 \pi^2 \epsilon} \; + \; 4 \lambda^2 \,\frac{3 \ICT (1+4\lambda^2 \JCT) - 11 \lambda \ICT^2 + 12\lambda^2 \ICT^3 - 11 \lambda \JCT}{1 - 8 \lambda \ICT + 12 \lambda^2 \ICT^2} \notag\\
&+ \ 6 \lambda h_t^4 \, \frac{-\ICT \Big(\frac{5}{16 \pi^2} + 6 \JCT \Big) + \frac{12 \lambda}{16 \pi^2} \ICT^2 + 4 \JCT}{1 - 8 \lambda \ICT + 12 \lambda^2 \ICT^2} \;,\displaybreak[0]\\
\delta \lambda_1^B \ &= \ - \, \frac{3\,h_t^4}{16 \pi^2 \epsilon}  \; + \; \frac{4 \lambda^2}{1 - 2 \lambda \ICT}\Big(3 \ICT - 5 \lambda \ICT^2 - 5 \lambda \JCT \Big) \; + \; \frac{6\lambda h_t^4 \JCT}{1 - 2 \lambda \ICT}\;,\displaybreak[0]\\
\delta \lambda_1^{\rm cb} \ &= \ \frac{\ICT}{1- 2 \lambda \ICT} \, \frac{6 \lambda h^4_t}{16 \pi^2} \;+\; \delta \lambda_{1, \mathrm{fin}}^{\rm cb} \;, \displaybreak[0]\\
\delta \lambda_2^A \ &= \ \frac{4 \lambda^2 \ICT}{1 - 8 \lambda \ICT + 12 \lambda^2 \ICT^2}\, (2 - 3 \lambda \ICT) \;,\displaybreak[0]\\
\delta \lambda_2^B \ &= \ \frac{2 \lambda^2 \ICT}{1 - 2 \lambda \ICT} \;.
\end{align}
\end{subequations}
Notice that we have added the finite contribution $\delta \lambda_{1, \mathrm{fin}}^{\rm cb}$ to $\delta \lambda_{1}^{\rm cb}$, which is chosen so as to ensure that $\Delta^{-1,G}(k=0;\, \phi=v)=\Delta^{-1,+}(k=0;\, \phi=v)=0$, as discussed in Section~\ref{sec:ferm}. In~case the two-loop self-energies $\bm{\Pi}^{\mathrm{2PI}, (2)}_a(k)$ and $\bm{\Sigma}^{\mathrm{2PI}, (2)}_a(k)$, as calculated in perturbation theory, are included in the EoMs given in~\eqref{eq:eomsren}, one then needs to add to the CTs~\eqref{eq:CTs} the relevant contributions resulting from standard perturbation theory. We shall not report their explicit expressions here. 

\newpage


\end{document}